\documentclass[useAMS,usenatbib]{mn2e}
\usepackage{graphicx,amssym}
\def\oii{[{O~\sc ii}]}
\def\nii{[{N~\sc ii}]}
\def\neiii{[{Ne~\sc iii}]}
\def\oiii{[{O~\sc iii}]}

\newcommand{\bc}{\begin{center}}
\newcommand{\ec}{\end{center}}
\newcommand{\cf}{\ifmmode C_f\else $C_f$\fi}

\title[The link between radio AGN activity and star formation]
      {Evolution of the Most Massive Galaxies to $z \sim $0.6: II. The link between radio AGN activity and star formation}
      
\author[Chen et al.]{
\parbox[t]{\textwidth}{\raggedright
Yan-Mei Chen$^{1}$\thanks{Email: chenym@nju.edu.cn},
Guinevere Kauffmann$^2$,
Timothy M. Heckman$^3$,
Christy A. Tremonti$^4$,
Simon White$^2$,
Hong Guo$^5$,
David Wake$^6$,
Donald P. Schneider$^{7,8}$
Kevin Schawinski$^9$
}\\
\vspace*{6pt}\\
$^1$Department of Astronomy, Nanjing University, Nanjing 210093, China\\
    Key Laboratory of Modern Astronomy and  Astrophysics (Nanjing University), Ministry of Education, Nanjing 210093, China\\
$^2$Max--Planck--Institut f\"ur Astrophysik,
    Karl--Schwarzschild--Str. 1, D-85748 Garching, Germany\\
$^3$Department of Physics and Astronomy, The Johns Hopkins University, 3400 North Charles Street, Baltimore, MD 21218\\
$^4$Department of Astronomy, University of Wisconsin-Madison, 1150 University Ave, Madison, WI 53706, USA\\
$^5$Department of Astronomy, Case Western Reserve University, 10900 Euclid Avenue, Cleveland, OH 44106\\
$^6$Astronomy Department, Yale University, New Haven, CT, 06520, USA\\
$^7$Department of Astronomy and Astrophysics, The Pennsylvania State University, University Park, PA 16802\\
$^8$Institute for Gravitation and the Cosmos, The Pennsylvania State University, University Park, PA 16802\\
$^9$Einstein Fellow, Yale Center for Astronomy \& Astrophysics, Yale University, New Haven CT 06520 U.S.A}

\begin{document}



\maketitle

\label{firstpage}

\begin{abstract}
We analyze the optical spectra of massive (log $M_*/M_\odot > 11.4$)
radio-loud galaxies at $z \sim$ 0.2 and $z \sim$ 0.6. Our samples are generated
by crossmatching the SDSS DR7 and BOSS spectroscopic galaxy catalogues
with the FIRST and NVSS radio continuum surveys. By comparing stellar
population parameters of these radio-loud samples with radio-quiet
control samples matched in stellar mass, velocity dispersion and
redshift, we investigate how the presence of a radio-emitting jet
relates to the recent star formation history of the host galaxy. We
also investigate how the emission-line properties of the radio
galaxies evolve with redshift by stacking their spectra. Our main
results are the following.  (1) Both at low and at high redshift, half
as many radio-loud as radio-quiet galaxies have experienced
significant star formation in the past Gyr. This difference in star
formation history is independent of the luminosity of the radio AGN,
except at radio luminosities greater than $10^{25.5}{\rm W~Hz}^{-1}$, where it
disappears. (2) The Balmer absorption line properties of massive
galaxies that have experienced recent star formation show that star
formation occurred as a burst in many of these systems. (3) Both the
radio and the emission-line luminosity of radio AGN evolve
significantly with redshift. The average \oiii\ rest equivalent width
increases by 1 dex from $z$ = 0.2 to $z$ = 0.6, and emission line ratios
change from LINER-like at low redshift to Seyfert-like at high
redshift. However, radio galaxies with similar stellar population
parameters, have similar emission-line properties both at high- and at
low-redshift.  These results suggest that massive galaxies experience
cyclical episodes of gas accretion, star formation and black hole
growth, followed by the production of a radio jet that shuts down further
activity. The behaviour of galaxies with log $M_*/M_\odot > 11.4$ is the same
at $z$ = 0.6 as it is at $z$ = 0.2, except that higher redshift galaxies
experience more star formation and black hole growth and produce more
luminous radio jets during each accretion cycle.
\end{abstract}


\section{Introduction}
\label{sec:intro}
Large redshifts surveys such as the Sloan Digital Sky Survey \citep[SDSS;][]{york00} and
the Two Degree Field (2DF) redshift survey \citep{colless01} have provided  
the sky coverage needed to define sufficiently large
samples of nearby radio sources to study the population statistics and global energetics of 
these systems at $z \sim 0.1$. \citet{best05b} investigated the properties of a sample of 2215
nearby radio galaxies with 1.4 GHz radio luminosity below $10^{25}{\rm W~Hz}^{-1}$ 
from SDSS, finding that the fraction of radio-loud AGN is a strong 
function of both black hole and stellar mass. \citet{mauch07} studied a sample of 2661 radio-loud 
AGN selected from the 6dF Galaxy Survey. They confirmed the findings of \citet{best05b} that 
radio-loud AGN preferentially inhabit the brightest and most massive host galaxies, and showed 
that the fraction of galaxies which host a radio-loud AGN correlates with the infrared $K$-band 
luminosity as $L_K^2$. \citet{best05b} also reported  that there was no correlation between radio 
luminosity and optical emission-line luminosity for the galaxies in their sample, concluding that
optical AGN and low-luminosity radio-loud AGN are independent phenomena which are triggered 
by different physical mechanisms. In later work, it was found that at fixed stellar mass, radio-loud 
AGN are preferentially located in denser environments than control samples of radio-quiet  
galaxies with similar masses and redshifts \citep{best07, wake08a, wake08b, mandelbaum09, donoso10} as well as control samples of 
optically-identified AGN \citep{kauffmann08, donoso10}. These results have led to the paradigm that 
strong  optical AGN activity is associated with galaxies with a significant cold gas reservoir and 
ongoing star formation \citep{heckman04}, while radio AGN activity may be associated with the 
accretion of hot gas in galaxies located at the centers of  massive dark matter 
halos \citep{bower06, croton06}.

Observational support for a scenario in which radio AGN play an important role in regulating the
growth of massive galaxies has accumulated rapidly over the past decade. X-ray studies of groups 
and clusters of galaxies with Chandra and {\it XMM-Newton} have shown that these jets interact strongly 
with their environment, blowing clear cavities or `bubbles' in the surrounding X-ray-emitting gas
\citep{bohringer93, fabian00, churazov01, fabian03, fabian05, forman05}. These cavities provide a 
direct estimate of the radio jet power \citep{birzan04, birzan08} and hence an empirical calibration of the 
relation between radio luminosity and mechanical power of the radio jet in nearby galaxies. The energy 
input from jets has been found to be sufficient to prevent star formation activity in massive early-type 
galaxies by heating the interstellar gas and suppressing the onset of cooling 
flows \citep{binney95, birzan04, rawlings04, best06, best07, schawinski06}.

If radio jets regulate ongoing star formation in massive galaxies, one might expect to see differences 
in the stellar populations of the host galaxies of radio-loud AGN compared to radio-quiet galaxies.  
\citet{best05b} compared the fraction of radio-loud AGN among galaxies with different values of the  
4000 \AA\ break strength (D4000), which is a powerful measure of the mean age of the stellar population 
in a galaxy. As can be seen from Figure~10 of that paper, there appears to be a lower fraction of 
radio-loud AGN among the most massive galaxies ($> 10^{11} M_\odot$) with
young stellar populations (i.e. low values of D4000). 
The statistics were rather poor, so Best et~al. did not place much emphasis on this result.  

While much of our knowledge about radio AGN is based upon studies of local galaxies, our understanding 
of the interplay between gas cooling processes and radio AGN feedback at higher redshifts is still very 
much in its infancy. Strong cosmological evolution of the high-luminosity radio source population has 
been long established  \citep[e.g.,][]{dunlop90}, but the role of this strongly evolving population in regulating 
the growth of massive galaxies at high redshifts is largely unconstrained. \citet{johnston08} constructed  
high signal-to-noise ratio (S/N) composite optical spectra of 375 radio-detected luminous red galaxies  at $0.4 < z < 0.8$. 
They compared composite spectra binned by radio luminosity and redshift  with those of radio-quiet 
control galaxies and concluded that there were no differences 
in the stellar populations of the two samples, except for the most powerful radio galaxies 
(with $L_{\rm 1.4GHz}  > 10^{26}{\rm W~Hz}^{-1}$), which had  stronger optical emission lines and  younger 
stellar populations. We note, however, that the galaxy survey  on which this analysis was based was  subject 
to a strong pre-selection on galaxy colour and would miss blue, star-forming galaxies 
enitirely. 

The Baryon Oscillation Spectroscopic Survey (BOSS; Dawson et al. 2012, submitted), one of 
the surveys of SDSS-III \citep{eisenstein11}, is currently obtaining spectra of 1.5 million luminous 
galaxies to $z \sim 0.7$. At redshifts greater than $z \sim 0.55$, the galaxy samples are reasonably complete
at stellar masses greater than $10^{11.4} M_{\odot}$. In this paper, we have cross-correlated galaxies with 
$z>0.55$ from the  BOSS spectroscopic survey with the National Radio Astronomy Observatory (NRAO) Very 
Large Array (VLA) Sky Survey \citep[NVSS;][]{condon98} and the Faint Images of the Radio Sky at Twenty 
centimeters (FIRST) survey \citep{becker95} and  have generated the largest sample of
intermediate redshift radio galaxies 
to date. We use this sample to investigate whether radio-loud AGN have had 
unusual star formation histories (SFHs) compared to galaxies of similar mass and redshift. Similar to \citet{johnston08}, 
we create composite optical spectra to study the emission line properties of radio-loud
and radio-quiet galaxies. By comparing 
results obtained for the BOSS samples with those from the low-redshift radio AGN catalogue of \citet{best12}, we study 
how the radio AGN population has evolved over a timescale of 3.3 Gyr.     

Our paper is arranged as follows. In \S2, we describe the low and high redshift radio-loud AGN samples and 
the construction of the radio-quiet control samples.  In \S3, we discuss how we                                  
extract information about the past star formation histories of the galaxies in our samples,
present results for the radio-loud AGN hosts and the control galaxies, and analyze
the emission line properties of radio galaxies using composite spectra . Our  results 
are summarized in \S4. We use the cosmological parameters $H_0=70~{\rm km~s^{-1}~Mpc^{-1}}$,
$\Omega_{\rm M}=0.3$ and $\Omega_{\Lambda}=0.7$ throughout this paper. 

\begin{figure*}
\bc
\hspace{-0.6cm}
\resizebox{17cm}{!}{\includegraphics{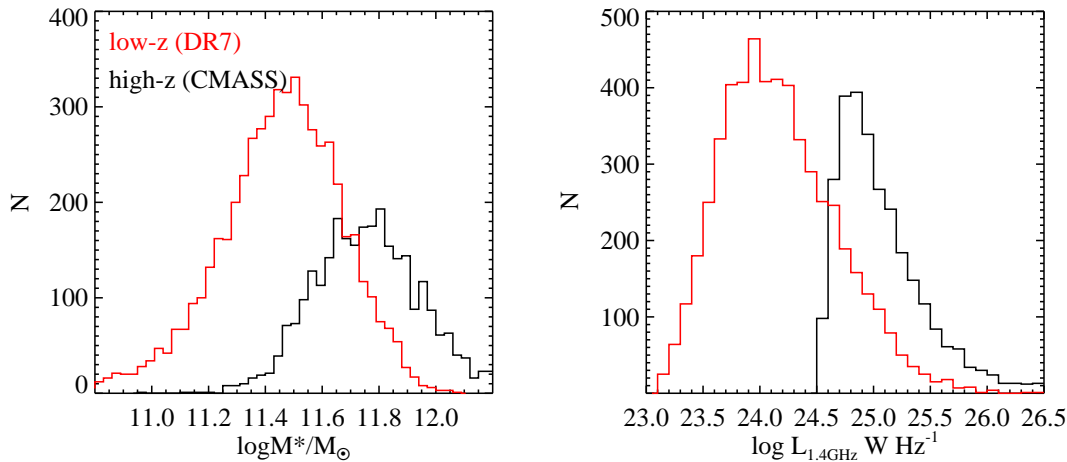}}\\%
\caption{The distribution of the stellar mass and radio luminosity for the DR7 (red) and CMASS (black) radio-loud AGN 
samples. Our DR7 sample includes 5818 galaxies with $0.1 <  z < 0.3$, while the CMASS sample
includes 2856 galaxies with $0.55 < z < 0.7$.
\label{mass_lradio_dist}}
\ec
\end{figure*}

\section{Sample selection}
\subsection{The low-redshift radio-loud AGN sample}
The Sloan Digital Sky Survey \citep[SDSS;][]{york00}  obtained photometry 
of nearly a quarter of the sky. A 
drift-scanning mosaic CCD camera \citep{gunn98} mounted on the SDSS 2.5m 
telescope at Apache Point Observatory \citep{gunn06} imaged the sky in the 
$u, g, r, i, z$ bands \citep{fukugita96}. The imaging data are astrometrically  
\citep{pier03} and  photometrically \citep{hogg01, pad08} calibrated  and used to 
select stars, galaxies, and quasars for follow-up fiber spectroscopy. 

The seventh data release \citep[DR7;][]{abazajian09} of the SDSS 
includes  about one million galaxy spectra. The spectra have a wavelength
coverage of $3800 - 9200$\AA\ and  are taken through 
$3^{\prime\prime}$ diameter fibers.
The low-redshift radio galaxy sample is described in  \citet{best12}. It has  a flux density limit of 5mJy, which 
means the sample extends  down to a radio luminosity of  $L_{\rm 1.4GHz}  \approx 10^{23}{\rm W~Hz}^{-1}$ 
at redshift $z = 0.1$. This sample was constructed by cross-correlating 927,552 galaxies from the DR7 MPA/JHU 
value-added spectroscopic catalogue\footnote{The MPA/JHU catalogue is available at http://www.mpa-garching.mpg.de/SDSS/.} 
with the NVSS and FIRST surveys. The cross-matching followed  the method described by \citet{best05a},  
but also included  the improvement developed by \citet{donoso09} for identification of sources 
without FIRST detections. \citet{best12} also separated radio AGN from star-forming galaxies. We refer the 
reader to their Appendix A for more details. 

For the current paper, we begin with a subset of 5818 radio galaxies with $r$-band magnitude $14.5 < r < 17.77$, 
redshift $0.1 < z < 0.3$ and SPECPRIMARY  flag = 1 drawn from \citet{best12} catalogue. The $r$-band magnitude 
limit restricts the sources to the `main' galaxy sample \citep{strauss02}. With the SPECPRIMARY  flag = 1, we avoid 
repeat observations of a given object. Properties of the radio source host galaxies required in this work include total stellar masses 
($M_*$), stellar velocity dispersions ($\sigma_*$) and SFH parameters derived from the principal component 
analysis method \citep[PCA,][]{chen12}. Radio luminosities are calculated using the formula
\begin{equation}
{\rm log}_{10}[L_{\rm 1.4GHz}] = {\rm log}_{10}[4\pi D_{\rm L}^2(z)S_{\rm 1.4GHz}(1+z)^{\alpha-1}]
\end{equation}
where $D_{\rm L}$ is the luminosity distance in the adopted cosmology, $S_{\rm 1.4GHz}$ is the measured radio 
flux density, $(1+z)^{\alpha-1}$ is the standard $k$-correction used in radio astronomy and $\alpha$ is the radio spectral 
index ($S_\nu \sim \nu^{-\alpha}$).  We adopted $\alpha=0.7$, as is usually assumed for FRI radio galaxies 
\citep{condon02}.
The red histograms in Figure~\ref{mass_lradio_dist} show the distribution of stellar mass and radio luminosities of the 
low redshift radio-loud AGN sample. Any galaxy that meets our radio flux cut is termed  ``radio-loud'' 
in this paper.

\subsection{The high-redshift radio-loud AGN sample}
The SDSS-III project has completed an additional 3000 ${\rm deg}^2$ of imaging 
in the southern Galactic cap in a manner identical to the original SDSS imaging \citep{aihara11}.
The Data Release 9 (DR9) of BOSS includes spectra of 462,979 luminous galaxies to  $z \sim 0.7$ 
(Ahn et~al. 2012, submitted). The spectrographs (Smee et al. 2012, submitted) have been  significantly upgraded from 
those used by SDSS-I/II, with improved CCDs with better red response, 
high throughput gratings. The new fibers are  $2^{\prime\prime}$ in diameter and the spectra
cover the  wavelength range  3600$-$10,000\AA. Galaxy spectroscopic redshift is determined
by the BOSS pipeline (Bolton et al. 2012, submitted).
The details of how targets are selected from the photometry are 
described in \citet{eisenstein11}.

The high redshift galaxy sample we analyze here is the ``CMASS'' sample (so-named, because it is very 
approximately stellar-mass limited). This sample is defined by the following cuts:
\begin{equation}
\begin{array}{c}
  d_\perp>0.55 \quad{\rm and}\quad
17.5 < i < 19.9 \quad {\rm and}\quad i_{\rm fibre2} < 21.5 \\
  i < 19.86+1.6\,(d_\perp-0.8) \quad {\rm and} \quad r-i<2
\end{array}
\label{eqn:cuts}
\end{equation}
where $d_\perp$ is a ``rotated'' combination of colours defined as $d_\perp=(r-i)-(g-r)/8$. The quantity $i_{\rm fibre2}$ is the 
$i$-band magnitude of the galaxy measured within a $2^{\prime\prime}$ BOSS fibre in the SDSS $ugriz$ 
photometric system \citep{fukugita96}. All magnitude limits are given in terms of ``cmodel" 
magnitudes whereas colour cuts are defined using ``model" magnitudes. Two additional conditions are 
introduced  to reduce  contamination by stars: $z_{\rm psf}-z_{\rm model}\ge 9.125-0.46\,z_{\rm model}$ 
and $i_{\rm psf}-i_{\rm model}> 0.2 + 0.2\times (20.0-i_{\rm model})$ (``psf'' refers to the $psfMag$ quantity 
in the SDSS database).  

The cuts listed in equation (2) are designed to identify massive galaxies at $z > 0.4$. The parameter $d_\perp > 0.55$ 
selects galaxies at high enough redshift that the 4000 \AA\ break has shifted beyond the observer frame 
$r$-band, leading to red observed $r-i$ colours. The cut on the $i$-band magnitude is designed to produce a 
sample that is approximately complete down to a limiting stellar mass of 
$M_* \sim 10^{11.2}M_\odot$ \citep{chen12}.  

We cross match the CMASS galaxies with NVSS and FIRST, using the same method as in \citet{best12}.   
We generate a sample of 5469 radio galaxies with a flux density limit of 3mJy. In this work, we limit the sample 
to 2856 galaxies with $z > 0.55$ where the colour incompleteness of the sample is minimized. As discussed in 
detail in \citet{chen12}, the $d_\perp > 0.55$  colour cut used to define the CMASS galaxies results in blue 
galaxies being missed from the survey, particularly at  lower redshifts. \citet{chen12} showed that the fraction 
of massive galaxies with recent star formation could be under-estimated by more than an order of magnitude 
at $z < 0.55$. At higher redshifts, the correction to the fraction of  actively star-forming galaxies is approximately 
a factor of two. The 3mJy flux limit corresponds to a radio luminosity of 
$L_{\rm 1.4GHz}  \approx 10^{24.5}{\rm W~Hz}^{-1}$ at redshift $z = 0.55$. At these very high radio luminosities, 
we will almost exclusively pick up radio-loud AGN rather than star-forming galaxies  \citep{sadler02, best05b}.

Once again, the properties of the CMASS radio source host galaxies,  including total stellar masses, stellar 
velocity dispersions  and SFH parameters, are derived from the principal component 
analysis method, as described in  \citep{chen12}. These authors found that the PCA method gives more robust 
estimations of D4000, H$\delta$A and $\sigma_*$ for the low S/N CMASS spectra 
($>$ 85\% galaxies have a median S/N per pixel (69$\rm km~s^{-1}$) below 4). The black histograms in Figure~\ref{mass_lradio_dist} 
show the distribution of stellar mass and radio luminosity of the high redshift radio-loud AGN sample. As can 
be seen, the high-redshift radio AGN are offset to higher stellar masses and radio luminosities compared to 
the low-redshift radio AGN, but there is a reasonable range in mass and luminosity where direct comparisons
between the two samples can be made.

\subsection{Radio-quiet control samples}
We create samples of radio-quiet galaxies,
which are matched as closely as possible to the radio-loud AGN hosts.
The purpose of the control samples is to determine which galaxy properties 
are causally linked with the radio-emitting jet.  

For each galaxy in the 
radio-loud sample, we find another galaxy located in the FIRST and NVSS survey area without a 
radio detection that is closely matched in redshift ($|\Delta z| < 0.005$), stellar mass 
($|\Delta {\rm log}M_*| < 0.05$), and velocity dispersion ($|\Delta \sigma_*| < 10 {\rm km~s}^{-1}$). 

The motivation for choosing this set of matching parameters is the following:
\begin{enumerate}
\item By constraining the control galaxy to be at similar redshift, we avoid any
confusion due to evolutionary effects 
and we make sure the SDSS fibre spectrum includes light coming from the the same physical 
scales in both the radio galaxy and the control.  
\item Constraining the control galaxies  to have similar stellar mass is extremely important because the  
probability that a galaxy is a radio source is a strong function of mass.  Stellar population properties 
are known to vary strongly as a function of galaxy mass.
\item Constraining the control galaxies  to have similar velocity dispersions ensures that the
radio-loud and the radio-quiet galaxy samples    
have roughly the same black hole mass distributions.  Any AGN physics that is determined  primarily by
the black hole mass will thus be the same for the two samples.
\end{enumerate}
Any radio galaxy without a radio-quiet twin is  excluded from our analysis. This leaves a total of  5416 (2820) 
radio galaxies in the DR7 (CMASS) samples. 

\begin{figure*}
\bc
\includegraphics[angle=0,width=1.0\textwidth]{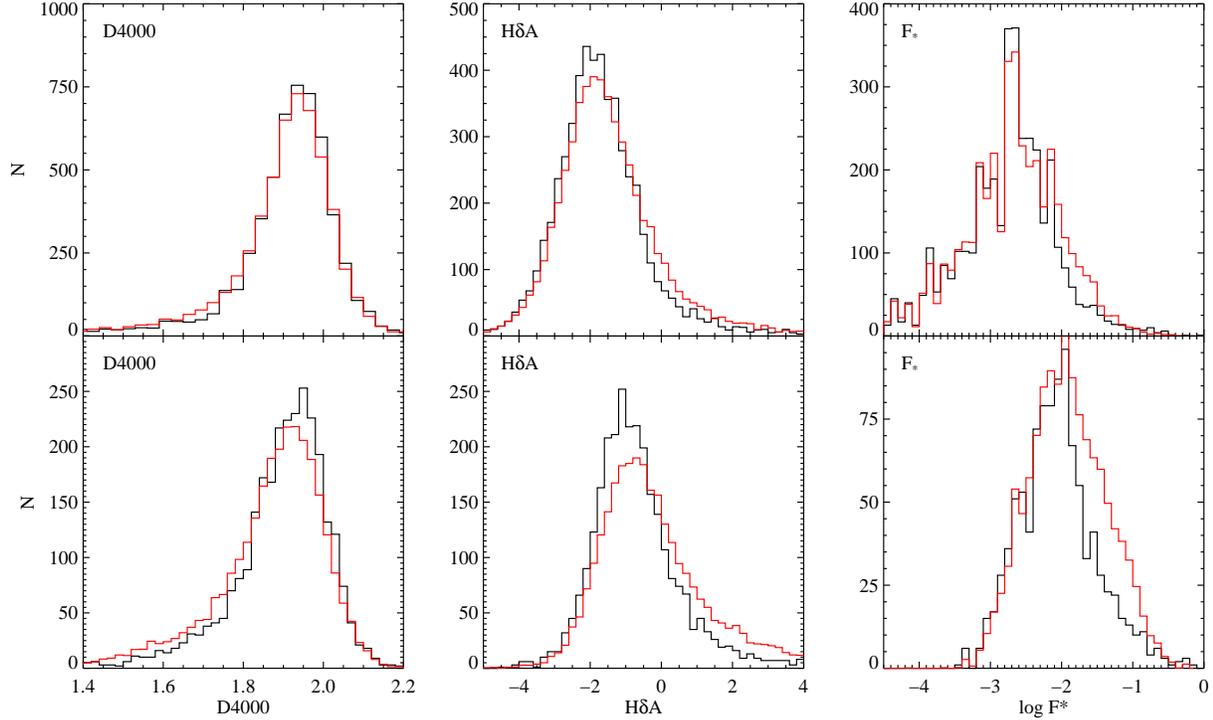}
\caption{Comparison of galaxy stellar population parameters D4000,  H$\delta$A and $F_*$ for 
the radio-loud galaxy samples and their radio-quiet control samples. The top panel shows results for  DR7; 
the bottom panel for CMASS. The radio-loud galaxies are shown as black histograms while the radio-quiet 
galaxies are shown as red histograms. The radio-loud  and radio-quiet samples  include exactly the 
same number of galaxies. There are 5416 and 2820 galaxies in the DR7 and CMASS samples, respectively.  
\label{dist_stellar_pop}}
\ec
\end{figure*}
\begin{figure*}
\bc
\hspace{-0.6cm}
\resizebox{17cm}{!}{\includegraphics{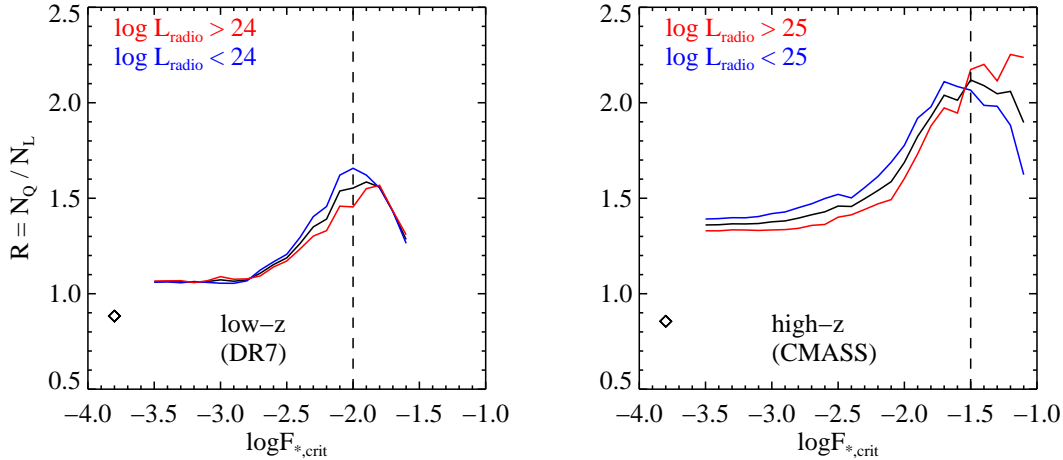}}\\%
\caption{The ratio $R$ between the number of galaxies with $F_* > F_{*, crit}$ in the radio-loud  and 
radio-quiet samples is plotted as a function of log$F_{*, crit}$. In the left panel, we show  results for DR7, 
while right panel is for CMASS. In each panel, the red line represents galaxies with $L_{\rm 1.4GHz}$ 
larger than the median radio luminosity ($10^{24}{\rm W~Hz}^{-1}$ for DR7, $10^{25}{\rm W~Hz}^{-1}$ 
for CMASS), while the blue line is for galaxies with $L_{\rm 1.4GHz}$ smaller than these values. The black line 
is the result for the whole sample. The diamonds in the bottom-left corner indicate the values of $R$ for 
galaxies with undetectable small $F_*$.  
\label{fstar_deltaf}}
\ec
\end{figure*}
\begin{figure*}
\bc
\hspace{-0.6cm}
\resizebox{17cm}{!}{\includegraphics{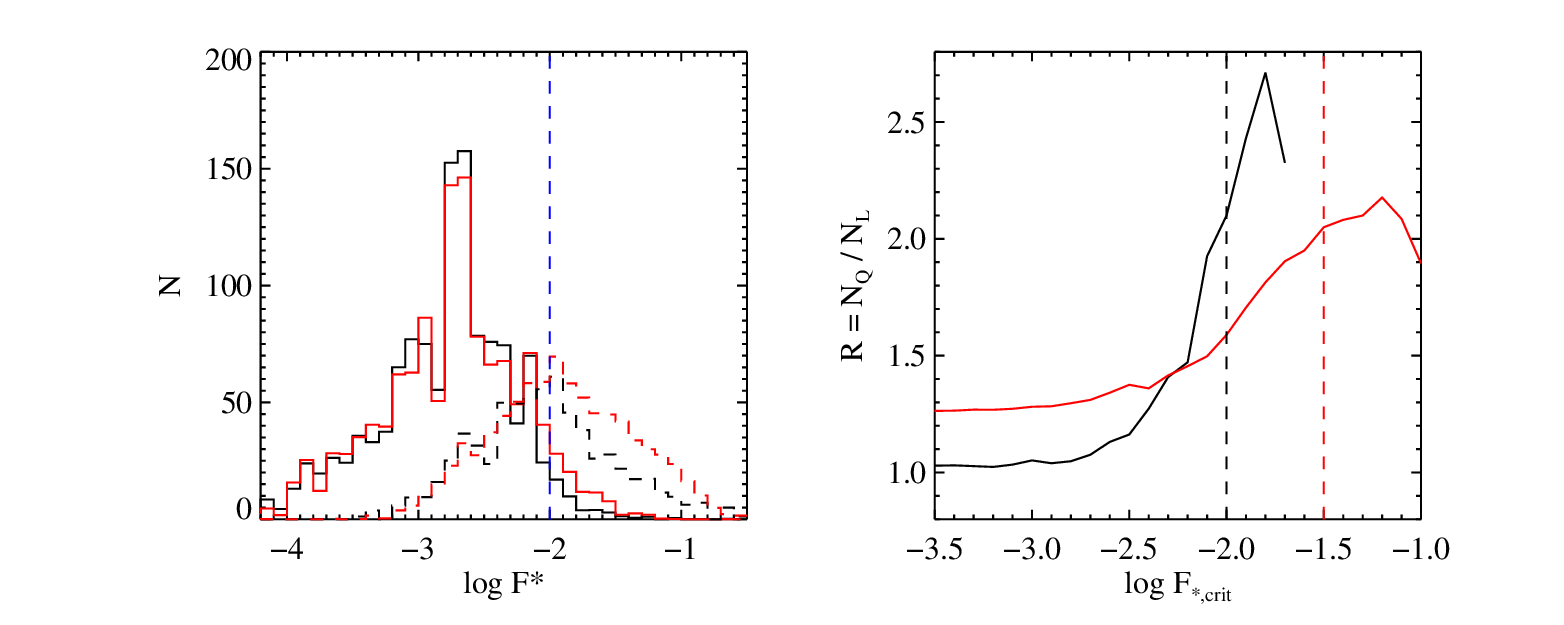}}\\%
\caption{We match the DR7 and CMASS radio-loud samples by stellar masses. Left panel:  the 
distributions of log$F_*$ for the mass-matched samples, solid lines for DR7, dashed lines  for CMASS,
black and red lines colour-code the radio-loud and radio-quiet galaxies, respectively. 
In both DR7 and CMASS, there are more actively star-forming galaxies with log$F_* > -2$ (the 
vertical blue line) in the radio-quiet sample than in the radio-loud sample. 
Similar to Figure~\ref{fstar_deltaf}, the right panel shows $R$ as a function of $F_{*, crit}$ for the DR7 (black) 
and CMASS (red) samples. 
\label{fstar_deltaf_two_z}}
\ec
\end{figure*}

\section{Results}
\subsection{Stellar populations of radio-loud and radio-quiet galaxies} 
In this section, we analyze the standard absorption line indices  D4000 and  H$\delta$A, reconstructed 
using our PCA technique, as well as the parameter $F_*$,  defined as the ratio between the mass of the stars 
formed in the last Gyr and the stellar mass of the galaxy. The quantity $F_*$ can be transformed into the star formation 
rate (SFR) averaged over the last Gyr using the equation,  SFR = $F_* \times M_*/10^9~M_\odot~{\rm yr}^{-1}$. 

Figure~\ref{dist_stellar_pop} presents histograms of these quantities for DR7 radio-loud galaxies and controls 
in the top panel, and for CMASS radio-loud galaxies and controls in the bottom panel. The radio-loud galaxy 
sample (black histograms) is clearly shifted to larger values of D4000, smaller values of H$\delta$A and smaller 
values of $F_*$\footnote {The number of galaxies included in the $F_*$ histograms is smaller than in the other 
two, because the majority of galaxies have undetectable small $F_*$.}. The largest differences are seen for 
H$\delta$A and for $F_*$ at both low and at high redshifts. 
We performed a Kolmogorov-Smirnov
(K-S) test on the D4000, H$\delta$A, $F_*$ distributions of radio-loud and radio-quiet samples and obtained a probability
smaller than $10^{-5}$ for these three parameters for the null assumption that the radio-loud and radio-quiet galaxies are
drawn from the same parent distribution.

We now focus on (1) quantifying the difference between the stellar populations of radio-loud and 
radio-quiet galaxies in more detail; (2) studying the evolution of this difference with  redshift. We 
define a parameter $R = N_{\rm Q}/N_{\rm L}$, where $N_{\rm L}$ ($N_{\rm Q}$) is the number 
of galaxies with $F_* > F_{*, crit}$ in the radio-loud (quiet) samples. $F_{*, crit}$ is a certain value of $F_*$. The left panel of 
Figure~\ref{fstar_deltaf} shows $R$ as a function of log$F_{*, crit}$ for DR7 galaxies,  while the 
right panel shows the same thing for CMASS galaxies. In each panel, the red line represents 
galaxies with $L_{\rm 1.4GHz}$ larger than the median radio luminosity ($10^{24}{\rm W~Hz}^{-1}$ 
for DR7, $10^{25}{\rm W~Hz}^{-1}$ for CMASS), while the blue line is for galaxies with 
$L_{\rm 1.4GHz}$ smaller than these values. The black line represents the whole sample. The diamonds 
in the bottom-left corner represent the ratio between the numbers of radio-quiet and radio-loud 
galaxies with $F_* \sim 0$. Figure~\ref{fstar_deltaf} indicates that, 1) the ratio $R$ is always larger than 
1, implying that the ratio of  star forming to passive galaxies is always higher in the radio-quiet sample, 
2) the ratio $R$ increases with $F_{*, crit}$, implying that radio AGN are more important for suppressing 
strong star formation as compared to weak star formation, 3) the differences in $R$ for low-luminosity and 
high-luminosity  radio galaxies are negligible. 

From Figure~\ref{fstar_deltaf}, one might be tempted to infer that  $R$ is larger at higher redshifts. 
However, it is not fair to compare the  DR7 and CMASS samples, because these two samples have 
very different stellar mass distributions (see the left panel of Figure~\ref{mass_lradio_dist}), and
the star formation history of a galaxy strongly depends on its stellar mass 
\citep[e.g.,][]{zheng07, chen09, karim11}. In order to study evolutionary effects, we match the DR7 
and CMASS galaxies by stellar mass. 
We bin both samples by stellar mass with $|\Delta {\rm log}M_*| < 0.02$. For each stellar mass bin, 
we count the number of galaxies in both DR7 ($n1$) and CMASS ($n2$) samples. If $n1 > n2$, we randomly
select $n2$ galaxies from this certain mass bin of DR7, otherwise, we select $n1$ galaxies from CMASS.
This matching process leaves 1658 galaxies in both DR7 and CMASS samples. 
Unless specified otherwise, all future plots are based on mass-matched samples.

The left panel of Figure~\ref{fstar_deltaf_two_z} shows the distributions of log$F_*$ for the four 
mass-matched  samples: solid lines show results for DR7, dashed lines  for CMASS; distributions for radio-loud 
and radio-quiet galaxies are shown as black and red histograms, respectively. Two conclusions are 
evident from this panel: (1) the distribution of $F_*$ is shifted to much higher values in the  CMASS 
sample than in the DR7 sample. As shown by \citet{chen12}, star formation in very massive 
galaxies ($M_* > 10^{11.4}M_\odot$) has evolved strongly since a redshift of $\sim$0.6; 
(2) in both DR7 and CMASS, there are more actively star-forming galaxies with log$F_* > -2$ (the 
vertical blue line) in the radio-quiet sample than in the radio-loud sample. The right panel 
shows $R$ as a function of $F_{*, crit}$ for the mass-matched  DR7 (black) and CMASS (red) 
samples. As can be seen,  the range in $R$ spanned by the two samples is much the same. For
both the DR7 and CMASS samples $R$ increases from $\sim 1$ at low $F_{*, crit}$ to a maximum 
value of around $2-2.5$.   The main difference is that the distribution of $F_*$ has been shifted 
to higher values in the CMASS sample, so the maximum occurs at a higher value of $F_{*, crit}$.

\begin{figure}
\bc
\hspace{-0.6cm}
\resizebox{8.5cm}{!}{\includegraphics{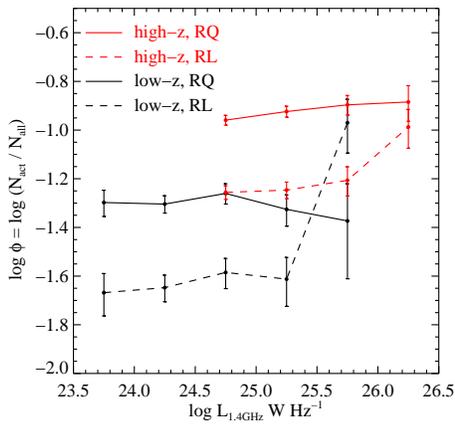}}\\%
\caption{This figure shows the fraction of actively star forming galaxies as a function radio luminosity. 
Actively star-forming galaxies are defined to be those with log$F_{*, crit} >  -1.5$ ($-2$) for the CMASS 
(DR7) samples. Black lines show results for DR7, while red lines are for CMASS. The dashed and solid 
lines represent the radio-loud galaxies and the radio-quiet controls, respectively. Note that for the 
radio-quiet galaxies, we use the radio luminosity from their radio-loud twins as the $x$-axis quantity.
RL and RQ represent radio-loud and radio-quiet respectively.
\label{f_vs_lradio}}
\ec
\end{figure}

We now ask whether the fraction  of actively star forming galaxies in the radio-loud sample depends 
on the threshold radio luminosity used to define it. We set log$F_{*, crit} = -1.5$ ($-2$) for the 
CMASS (DR7) samples and we define the parameter $\phi = N_{\rm act}/N_{\rm all}$ to represent 
the fraction of galaxies with $F_* > F_{*, crit}$. Figure~\ref{f_vs_lradio} shows $\phi$ as a function radio 
luminosity. Black lines are for DR7, while red lines are for CMASS. The dashed and solid lines 
represent radio-loud and radio-quiet galaxies, respectively. As expected, the fraction of actively 
star forming galaxies is higher in the radio-quiet samples. The main point of this plot is that  
$\phi$ remains  constant over a wide range in radio luminosity. At the very highest radio luminosities 
($L_{\rm 1.4GHz}  > 10^{25.5}{\rm W~Hz}^{-1}$ for DR7, 
$L_{\rm 1.4GHz}  > 10^{26}{\rm W~Hz}^{-1}$ for CMASS),  $\phi$ suddenly increases and the difference 
in the fraction of actively star forming galaxies in the radio-loud and the
radio-quiet samples  disappears. 

\subsection{Constraints on the star formation histories of massive galaxies}
Figure~\ref{dist_stellar_pop} shows that the difference in H$\delta$A between radio-loud AGN and radio-quiet 
control galaxies is much larger than the difference in D4000. This result is also true for the mass-matched samples. 
As discussed in \citet{kauffmann03a, kauffmann03b}, the location of galaxies in the D4000$-$H$\delta$A plane 
can tell us whether they have been forming stars in a burst or continuously over the last $1-2$ Gyr.
Galaxies with continuous star formation occupy a narrow strip in the D4000$-$H$\delta$A plane, while a recent 
burst displaces them away from this locus to higher values of H$\delta$A  for $\sim 1-2$ Gyr following the burst.

\begin{figure*}
\bc
\hspace{-0.6cm}
\resizebox{17cm}{!}{\includegraphics{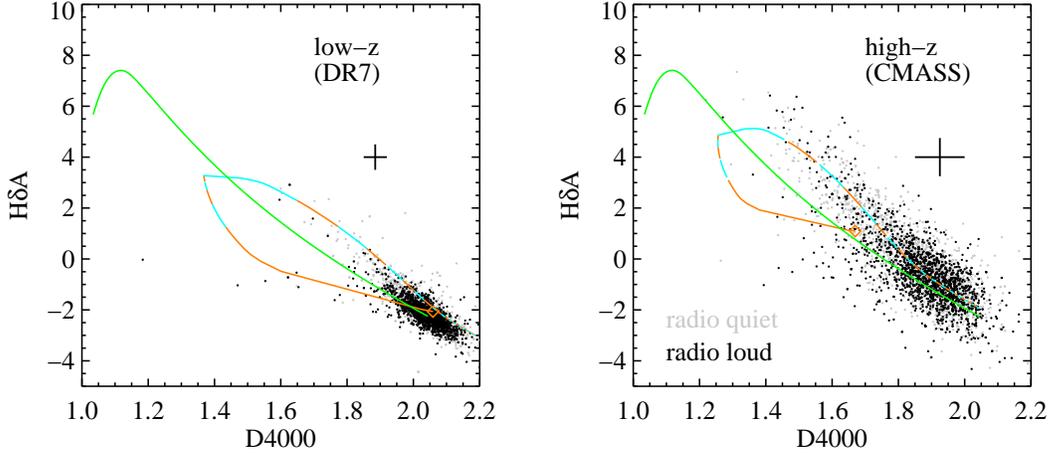}}\\%
\caption{The DR7 (left) and CMASS (right) galaxies on the D4000$-$H$\delta$A plane. Black dots represent 
radio-loud galaxies and grey dots represent radio-quiet galaxies. The over-plotted green line shows the locus 
of a model galaxy with exponentially declining SFH with time constant  $\tau=2$ Gyr and twice solar metallicity.  
The  orange-cyan line shows how a galaxy with an underlying old stellar population that experiences a recent 
burst of star formation evolves in the D4000$-$H$\delta$A plane. This model is calculated using \citet{bc03} 
population synthesis templates. The SFH consists of two parts: an underlying continuous model with $\tau = 1$ 
Gyr plus a recent burst. The burst, which has twice solar metallicity and duration 0.5 Gyr, occurs when the galaxy 
is 8 (5) Gyr old and the fraction of stellar mass produced during the burst relative to the total mass formed by the 
continuous model is $\sim$1\% ($\sim$3\%) in the left (right) panels. The orange diamonds mark the 
place where the burst begins. Following the burst, every colour-coded  section of the track represents a duration 
of 100 Myr. The median $\pm1\sigma$ measurement errors on D4000 and H$\delta$A are indicated in the right 
side of each panel.
\label{d4nhda}}
\ec
\end{figure*}

Figure~\ref{d4nhda} shows the DR7 (left) and CMASS (right) galaxies in the D4000$-$H$\delta$A plane. Black 
dots represent radio-loud AGN  and grey dots represent radio-quiet galaxies. The over-plotted green line shows the locus 
of a model galaxy with exponentially declining SFH with time constant  $\tau=2$ Gyr and twice solar metallicity.  
The  orange-cyan line shows how a galaxy with an underlying old stellar population that experiences a recent 
burst of star formation evolves in the D4000$-$H$\delta$A plane. This model is calculated using \citet{bc03} 
population synthesis templates. The SFH consists of two parts: an underlying continuous model with $\tau = 1$ 
Gyr plus a recent burst. The burst, which has twice solar metallicity and duration 0.5 Gyr, occurs when the galaxy 
is 8 (5) Gyr old and the fraction of stellar mass produced during the burst relative to the total mass formed by the 
continuous model is $\sim$1\% ($\sim$3\%) in the left (right) panels. In the plot, the orange diamonds mark the 
place where the burst begins. Following the burst, every colour-coded  section of the track represents a duration 
of 100 Myr. 

These models are meant to be purely illustrative, and the values of burst duration and mass fraction 
that we have adopted are not tuned to fit to the data. The main conclusion that one reaches from studying 
Figure~\ref{d4nhda} is that the majority of CMASS and DR7 galaxies with D4000 $< 1.7$ lie systematically 
above the locus occupied by galaxies that have experienced exponentially declining star formation histories.
These galaxies have D4000/H$\delta$A values that can be explained if they have experienced recent 
bursts of star formation.

\begin{figure}
\bc
\hspace{-0.6cm}
\resizebox{8.5cm}{!}{\includegraphics{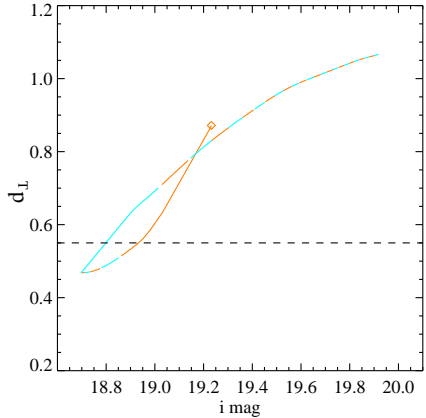}}\\%
\caption{The predicted behaviour of the colours of the burst model in the 
right panel of Figure~\ref{d4nhda} with $M_* = 10^{11.5}M_\odot$ 
at $z=0.6$. This figure shows the locus of this burst model in the $d_\perp - i$ plane, 
where $i$ represents the $i$-band magnitude. The orange diamond marks the point 
when burst begins.  Following the burst, every colour-coded part of the locus represents a duration of 100Myr. 
The horizontal dashed line marks the CMASS $d_\perp$ cut.
\label{target_sel}}
\ec
\end{figure}

There are many more galaxies in the post-burst phase in which H$\delta$A is offset to large 
values than  in the ``young burst phase'', when H$\delta$A is low (in fact, often in emission).  Part of the 
reason for this behaviour is that the time spent in the young burst phase is short. In addition, we caution that the $d_\perp$ 
colour cut used to define CMASS sample may exclude galaxies with a significant population of very young stars. 
Figure~\ref{target_sel} shows the track of the  burst model in the right panel of Figure~\ref{d4nhda} in the 
$d_\perp - i$ plane, where $i$ is  the $i$-band magnitude of the galaxy. A galaxy with a stellar mass of 
$10^{11.5} M_\odot$ at the end of the burst is assumed. Once again, the orange diamond marks the point 
when burst begins; following the burst, every colour-coded part of the locus represents a duration of 100Myr. 
The horizontal dashed line marks the $d_\perp$ cut. It is clear that a $3 \%$  burst can result in blue enough 
colours that a fraction of galaxies at $z=0.6$ may drop out of the sample. The burst track plotted in
Figure~\ref{target_sel} assumes  no observational errors, which will scatter some young bursts into the region of 
$d_\perp > 0.55$. The effect of dust on the colours has also not been included in the models.

\begin{figure*}
\bc
\hspace{-0.6cm}
\resizebox{17.cm}{!}{\includegraphics{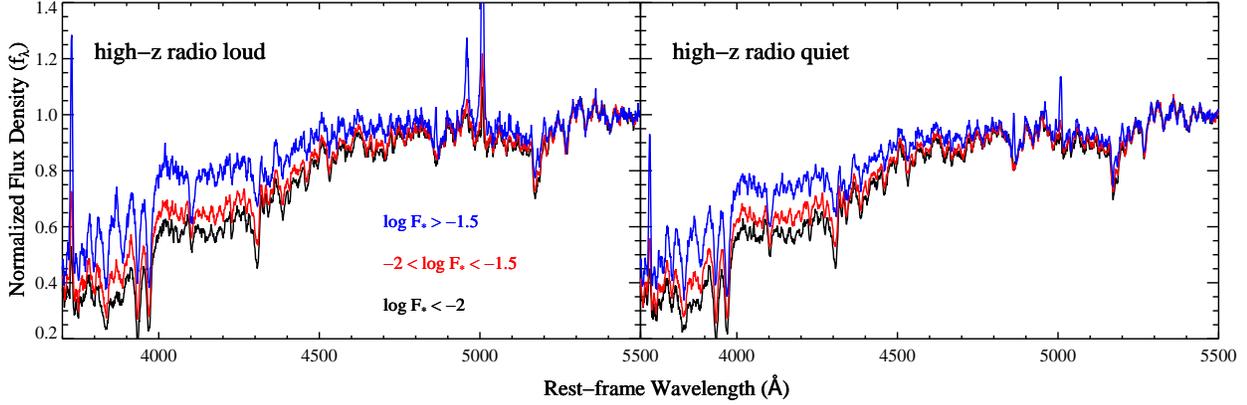}}\\%
\caption{Composite spectra of CMASS galaxies in different $F_*$ bins. The left panel shows results for  radio-loud 
galaxies and the right panel for radio-quiet galaxies. RL and RQ represent radio-loud and radio-quiet respectively.
\label{stackfs}}
\ec
\end{figure*}
\begin{figure*}
\bc
\hspace{-0.6cm}
\resizebox{17.cm}{!}{\includegraphics{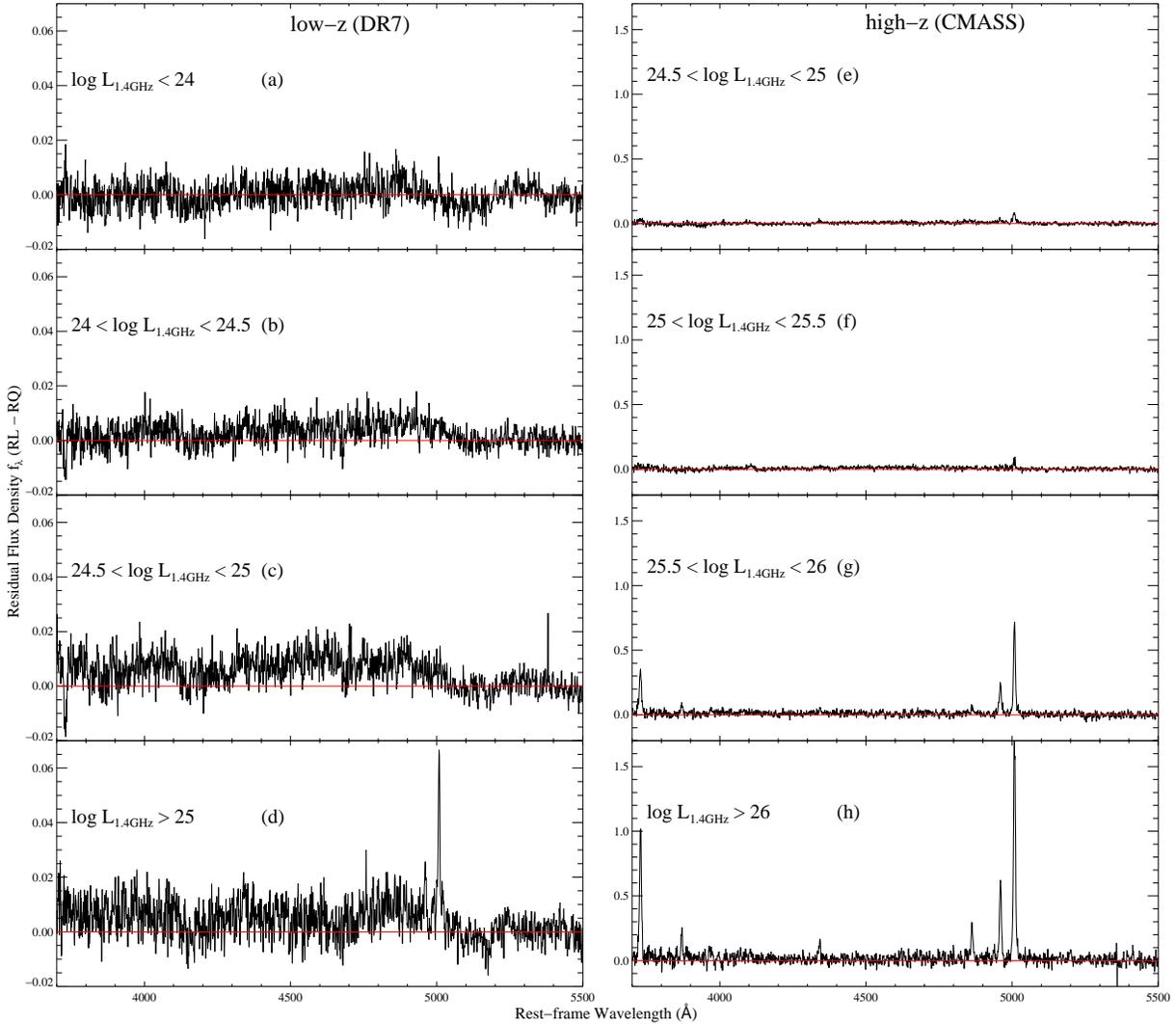}}\\%
\caption{The {\em difference} between the composite spectra of radio-loud and radio-quiet galaxies for the different 
radio luminosity bins. Radio luminosity increases from top to bottom. The left panel shows results for DR7 and the right 
panel for CMASS. The red lines mark the zero point (i.e. no difference between the spectra).
\label{stack_spec_lradio}}
\ec
\end{figure*}
\begin{figure*}
\bc
\hspace{-0.6cm}
\resizebox{17.cm}{!}{\includegraphics{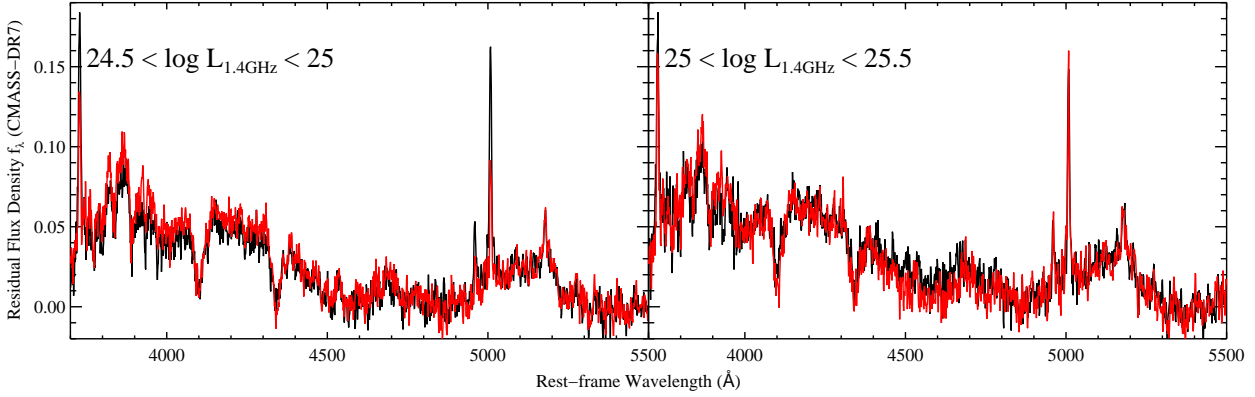}}\\%
\caption{The difference between CMASS (high-$z$) and DR7 (low-$z$) composites. Black lines show results for radio-loud 
galaxies, while red lines show results for radio-quiet galaxies. It is clear that the composites of CMASS galaxies are 
bluer and have stronger Blamer absorption lines than the DR7 galaxies although they have similar stellar masses.
\label{stack_spec_lradio1}}
\ec
\end{figure*}
\begin{figure*}
\bc
\hspace{-0.6cm}
\resizebox{17.cm}{!}{\includegraphics{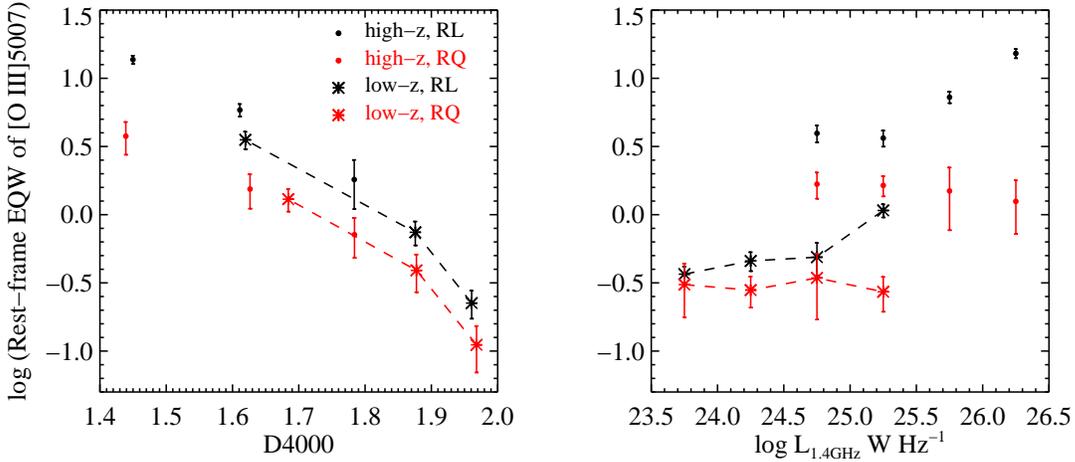}}\\%
\caption{This figure shows the correlations between 
\oiii\ EQW and D4000 (left), radio power (right) for the composites in various bins of D4000 and radio power. 
Black symbols represent radio-loud galaxies and red symbols represent radio-quiet galaxies. Stars connected 
by dashed lines represent galaxies from  DR7 and solid points represent galaxies from CMASS. 
We estimate the sample variance errors using a bootstrap technique. RL and RQ represent radio-loud and radio-quiet respectively.
\label{pop_vs_o3}}
\ec
\end{figure*}
\begin{figure*}
\bc
\hspace{-0.6cm}
\resizebox{17.cm}{!}{\includegraphics{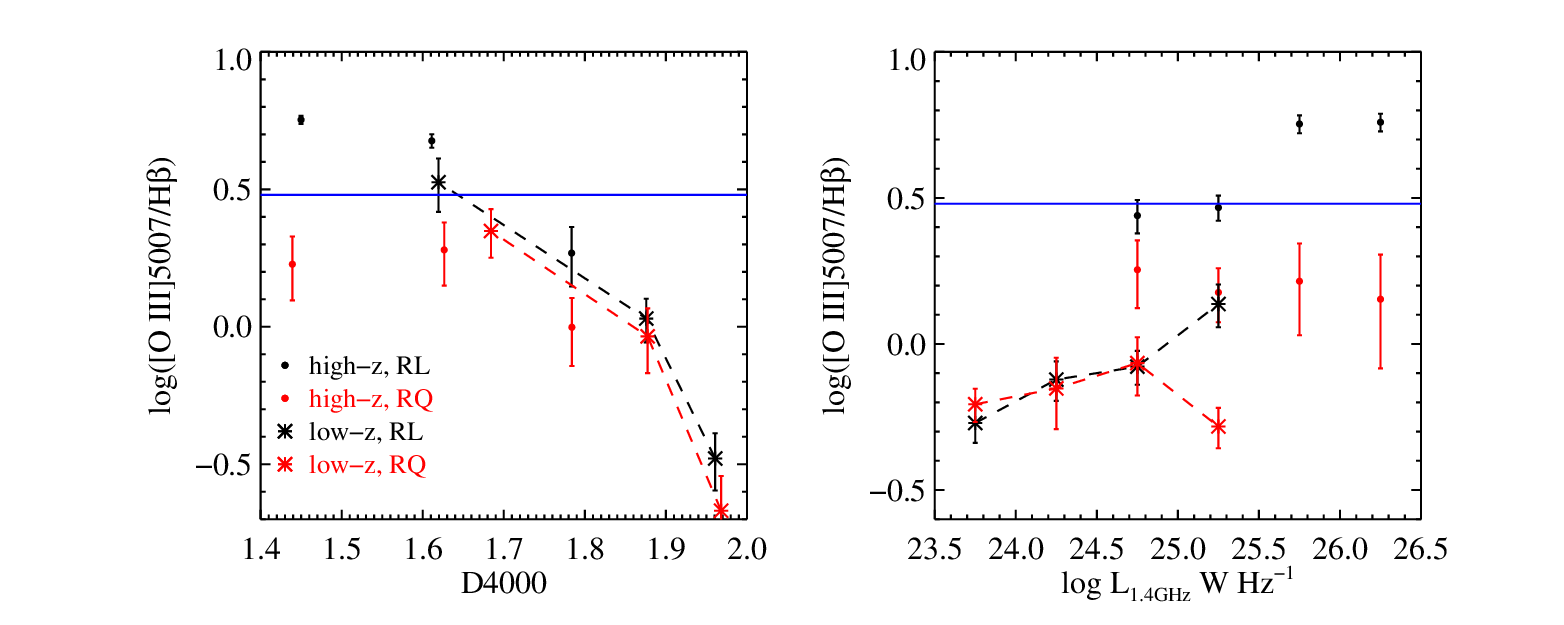}}\\%
\caption{The line ratio between \oiii\ and H$\beta$ as a function of D4000 (left panel) and radio power (right panel), 
the symbols and lines have the same meaning as that in Figure~\ref{pop_vs_o3}. The blue horizontal line 
demarcates the value of \oiii/H$\beta$ that separates Seyferts from LINERs in the AGN population 
\citep{kauffmann03c}. We estimate the sample variance errors using a bootstrap technique.
RL and RQ represent radio-loud and radio-quiet respectively.
\label{pop_vs_o3hb}}
\ec
\end{figure*}

\subsection{Emission line properties from composite spectra}
\subsubsection{Stacking by stellar population age} 
In this section, we analyze composite spectra in bins of $F_*$,  D4000 and H$\delta$A for the 
DR7 and CMASS samples. The galaxy spectra are first corrected for foreground Galactic attenuation 
using the dust maps of \citet{schlegel98}, transformed from flux densities to luminosity densities, and 
shifted to the rest frame using the redshift determined by the SDSS DR7/BOSS pipeline. Since most 
of the galaxies in our samples are intrinsically luminous galaxies, we normalize each spectrum by its 
mean luminosity in the wavelength range 5300$-$5500\AA, and simply average the spectra with equal 
weight (note that the weight of the bad pixels identified in the SDSS mask array is set to zero). 

Figure~\ref{stackfs} shows normalized composites of CMASS galaxies with log$F_* < -2$ (black), 
$-2 <$ log$F_* < -1.5$ (red) and log$F_* > -1.5$ (blue). We have not attempted to divide the galaxies 
with log$F_* < -2$ into separate bins because the PCA method is not able to recover the SFHs of galaxies
in which less than $\sim$1\% of the stellar population formed in the last Gyr. As expected, for both 
radio-loud (left panel) and radio-quiet (right panel) galaxies, emission line fluxes (e.g. \oii, \oiii) increase 
for larger values of $F_*$ and the spectra also become bluer, and the bluest galaxies have stronger emission 
lines if they are radio-loud than radio-quiet. Composites binned by D4000 and 
H$\delta$A yield very similar results, so we do not present them here.   

\subsubsection{Stacking by radio luminosity}
Figure~\ref{stack_spec_lradio}  shows the {\em difference} between the composite spectra of radio-loud 
and radio-quiet galaxies for different bins in radio luminosity. Radio luminosity increases from top to bottom 
in the plot. The left panel shows results for DR7 and the right panel for CMASS. The red lines mark zero flux density difference. 
The near zero residual flux in Figure~\ref{stack_spec_lradio}, except for regions around the emission lines, 
indicates that the {\em average} stellar continua of radio-loud and radio-quiet galaxies are quite similar. 
At first sight, this result may appear to conflict with the conclusions presented in the previous section. However, 
when we check the \citet{bc03} stellar population models, we find that a simple stellar population with an 
age of 0.5 Gyr is typically 10 times less luminous in the $r$-band than a 5 Gyr old stellar population of  similar 
mass. In the CMASS radio-quiet sample, the fraction of galaxies that have formed more than $\sim$3\% of their 
stellar mass in the last Gyr is about 10\%. We estimate that  if we stack all the CMASS radio-quiet galaxies
together, the fraction of light contributed by stars less than 1 Gyr in age
will contribute only  $\sim$ 3\% of the total flux in the composite 
spectrum. This explains why the stellar continua of the radio-loud and radio-quiet composite spectra are 
nearly identical.

There are almost no emission lines visible in the  composite spectra of the DR7 galaxies, except  weak 
\oiii$\lambda\lambda$4959,5007, which appears in panel (d) in the
composite spectra of  the most powerful radio sources.   
Almost all very massive galaxies ($M_* > 10^{11.4}M_\odot$) are inactive at low redshifts. 
In contrast, the 
spectra of the CMASS radio-loud galaxies have obvious emission lines (e.g., \oii, \neiii, H$\beta$, \oiii) 
and the emission line strengths increase strongly with  radio power. 

There are two bins of  radio luminosity  where we can compare the DR7 and CMASS composite spectra 
directly. Figure~\ref{stack_spec_lradio1} shows the residual between CMASS and DR7 composites for 
these two bins: black lines are for radio-loud galaxies, while red lines are for radio-quiet galaxies. It is 
clear that the CMASS composite spectra are bluer and have stronger Balmer absorption lines.
This indicates  
that the high redshift massive galaxies have significantly younger stellar populations on average. 

\subsubsection{Correlations between emission line properties and star formation}
Following the methodology described in \citet{tremonti04} and \citet{brinchmann04}, we fit a linear 
superposition of stellar population templates  to the stellar continua of the composite spectra presented 
in \S3.3.1 \& 3.3.2. We subtract the continuum model from the composite, and measure the rest equivalent width 
(EQW) of \oiii$\lambda$5007 from the residual spectrum.

Correlations between \oiii\ EQW and D4000 and radio power are shown in Figure~\ref{pop_vs_o3}. Black 
and red symbols represent radio-loud galaxies and radio-quiet galaxies, respectively. Stars connected 
by dashed lines represent galaxies from  DR7 and solid points represent galaxies from CMASS. As can 
be seen from the left panel, the strength of \oiii\ emission correlates strongly with D4000 for both 
radio-quiet galaxies and for radio-loud AGN. In the DR7 galaxy sample, the emission lines from
extremely massive galaxies ($> 10^{11.4} M_\odot$) almost always have ratios indicative of ionization by
``hard spectrum'' sources \citep[AGN or post-asymptotic giant branch stars;][]{kauffmann03c}. 
For the CMASS galaxies, \nii\ and H$\alpha$ have shifted out of the observed spectral range, but
alternative diagnostic diagrams such as EQW-\oii/EQW-H$\beta$ vs. log EQW-\oii\ \citep{rola97} 
indicate that the emission is still dominated by AGN.

If we regard the \oiii\ line as a tracer of the accretion of matter onto a central black hole\footnote {Population 
synthesis models (e.g. Bruzual \& Charlot 2003) show that [OIII] EQW of more than a few \AA\ cannot be 
produced by PAGB star ionization.},  the left panel of Figure~\ref{pop_vs_o3} indicates  that  
the average black hole growth  is correlated with the growth of the galaxy,  
as has been discussed previously  \citep{heckman04}. It is interesting
that at fixed D4000, radio AGN 
have somewhat higher \oiii\ EQW values than the control galaxies. This  suggests that shock excitation 
from the jet may contribute to the \oiii\ emission. The effect is quite weak (less than a 50\% change
in \oiii\ EQW). To zeroth order, one  relation
between \oiii\ EQW and  D4000 seems to adequately describe radio-loud and radio-quiet galaxies
at  $z \sim 0.2$ and at $z \sim 0.6$.  

The right panel of Figure~\ref{pop_vs_o3} shows trends in  \oiii\ EQW as a function of radio power
for the composite spectra constructed from the DR7 and CMASS samples. This plot
shows that 
the \oiii\ EQW 
does not correlate with radio power at radio luminosities below $\sim 10^{25.5}$ W Hz$^{-1}$ 
both for the DR7 and the CMASS radio samples.
At {\em fixed radio luminosity}, 
high-redshift radio galaxies have higher \oiii\ EQW than low-redshift radio galaxies. The same 
effect is seen in the control sample, so  
we attribute this evolution to the fact the   
amount of recent star formation in the massive galaxy population increases with 
redshift.
At radio luminosities higher than $\sim 10^{25.5}$ W Hz$^{-1}$, 
the \oiii\ EQW increases as a function of radio power. This is {\em not seen in the control sample 
galaxies}, so this phenomenon must be intrinsic to the radio-emitting jet itself.   

Figure~\ref{pop_vs_o3hb} shows the ratio between the \oiii\ and H$\beta$ line fluxes as a function 
of D4000 and radio power. The symbols and lines have the same meaning as that in 
Figure~\ref{pop_vs_o3}. The blue horizontal line demarcates the value of \oiii/H$\beta$
that separates Seyferts from LINERs in the AGN population \citep{kauffmann03c}. 
The trends visible in this plot mirror what is seen in Figure~\ref{pop_vs_o3}.
There is a strong correlation between  \oiii/H$\beta$ and D4000, which does not depend on
redshift and which is roughly the same for radio-loud and radio-quiet galaxies (we again 
see a slight excess in \oiii\ emission for the radio-loud population in this plot; it is 
particularly evident at low values of D4000). The ``universal''  correlation between \oiii/H$\beta$ 
and D4000  again leads us to attribute much of the evolution in  \oiii/H$\beta$  in the right 
panel of Figure~\ref{pop_vs_o3hb} to the changing stellar populations of high-redshift galaxies. 
However, the abrupt increase in \oiii/H$\beta$ to Seyfert-like values found at radio luminosities
in excess of $\sim 10^{25.5}$ W Hz$^{-1}$ is not seen in control samples, and must
be intrinsic to the most powerful radio sources.
 
Recall that in the previous subsection, we found that stellar population differences
between radio-loud galaxies and the radio-quiet controls disappears at radio
luminosities above $\sim 10^{25.5}$ W Hz$^{-1}$. These highly luminous radio galaxies
either correspond to a very early phase of the evolution of the black hole/jet before
star formation has been suppressed, or they represent a separate population of black 
holes that are undergoing a different accretion mechanism. We will investigate these 
questions in more depth in future work.

\section{Summary}
We have cross-matched the SDSS-III BOSS galaxy spectroscopic sample with the NVSS 
and FIRST surveys, generating the largest radio galaxy sample with
optical spectroscopy at $z\sim0.6$ to date.
Combining this sample with the SDSS DR7 radio-loud AGN catalogue of  \citet{best12}, 
we study the evolution of the recent star formation histories of the host galaxies of
radio-loud AGN, as well as  radio-quiet ``control'' galaxies that are matched in 
redshift, stellar mass, and velocity dispersion.  

In previous work (Chen et al 2012), we found that the fraction of massive galaxies that
have formed a significant fraction ($>$ a few  percent) of their stars in the past Gyr
has  strongly decreased with redshift from $z \sim 0.6$ to the present day.
The goal of this paper is to analyze if this evolution is linked to the radio-emitting jets 
hosted by massive galaxies. Our main empirical results may be summarized as follows:

\begin{enumerate}
\item  For galaxies with stellar masses $M_* > 10^{11.4}M_\odot$, the fraction of actively 
star-forming galaxies is $\sim 2$ times lower among galaxies with radio-emitting jets
than in the  radio-quiet control samples. This is true  both at low- and at high- redshift. 
The factor of two difference in the fraction of star-forming galaxies is independent of  radio luminosity, 
except for radio galaxies with luminosities in excess of $10^{25.5}$ W Hz$^{-1}$, 
where the difference disappears.   

\item The locus of massive galaxies in the D4000$-$H$\delta$A plane suggests that 
their star formation histories are characterized by bursts rather than low-level continuous 
star formation.  A smaller fraction of radio-loud galaxies have undergone significant ($>$ a 
few percent in mass fraction) starbursts in the last $\sim$Gyr compared to radio-quiet objects. 

\item There is a strong correlation between \oiii\ EQW and stellar population age (D4000) both 
for radio-loud AGN and for the control sample of radio-quiet galaxies.  
The relation between \oiii\ EQW and D4000 does 
not change with redshift.

\item
At fixed radio power, there is strong positive evolution in \oiii\ EQW with redshift.
The \oiii/H$\beta$ ratio also increases at higher redshift, i.e.; radio galaxies
are more LINER-like at low redshift and more Seyfert-like at high redshift.  
We interpret these correlations as {\em induced trends}, which follow from the fact that
all high redshift massive galaxies have more gas,  younger stellar populations and
larger black hole accretion rates.

\end{enumerate}

Our results suggest a picture in which massive galaxies experience cyclical episodes of
gas accretion, star formation and black hole growth, followed by production of a radio 
jet that acts to shut down the star formation in the galaxy. The behaviour of galaxies with
$M_* > 10^{11.4}M_\odot$ is the same at $z=0.6$ as it is at $z=0.2$, except that
higher redshift galaxies experience more black hole growth and star formation
and produce more luminous radio jets during each accretion cycle.

Let us imagine a scenario in which gas is gradually accreting onto a massive galaxy.
There may be an ``accumulation phase" where the cold gas gradually builds up
in a disk  with little or no star formation. After some critical gas density is reached, the disk
may become unstable and lose angular momentum \citep{salim12}. Molecular clouds form and star formation 
is triggered. This corresponds to the  ``burst phase". The radio source is triggered when 
gas finally reaches the central black hole. As the jet expands, it pushes away the surrounding cold gas,
causing the  starburst to shut down. The radio lobes expand into the surrounding intra-cluster 
medium, heat the ambient gas and shut down cooling onto the central galaxy. The shutdown 
of cooling removes the power source of the jet and the galaxy eventually returns to its original 
radio-quiet accumulation phase.

This picture is consistent with a population of radio galaxies with lower cold gas content and 
older stellar populations than radio-quiet galaxies. The strong evolution in the fraction of massive galaxies with 
significant star formation may  be explained by increased cooling efficiencies in massive halos 
at higher redshift \citep[see Figure~2 of ][]{white91}. If the accumulation and starburst phases 
have roughly equal durations, this would lead to a 2:1 ratio of starbursts in the radio-quiet and 
radio-loud populations, as observed. Recall, however, that clustering analyses demonstrate that 
radio-loud AGN are preferentially located at the centers of massive dark matter halos 
\citep{best07, wake08a, wake08b, mandelbaum09, donoso10}  where gas cooling and accumulation rates are presumably higher. A fair 
assessment of the jet/burst duty cycle requires one to control for such environment-dependent effects. However, in CMASS samples,
most galaxies are classified as central galaxies, satellites are a very small fraction of the total $-$ this is because of the very high 
stellar mass selection (Kovac et al. 2012, in preparation).

The missing link in our analysis of the accretion/starburst/jet cycle in massive galaxies 
is direct observations of the gas 
itself. An analysis of Chandra X-ray Observatory archival data of the hot gas in 222 nearby galaxy 
clusters revealed that H$\alpha$ and radio emission from the brightest cluster galaxy is much 
more pronounced when the the cluster's core gas entropy is low \citep{cavagnolo08}. Systematic 
studies of atomic and molecular gas in massive galaxies are still limited to small samples 
\citep[e.g.][]{schawinski09, catinella10, saintonge11}. Because radio galaxies are rare, large wide-field 
surveys will be required before samples are large enough to study how the cold interstellar medium 
of massive radio-loud galaxies differs from that of their radio-quiet counterparts. In the shorter term, 
it will  possible to stack DR7 and BOSS spectra to search for signatures of ionized gas outflows 
in luminous  radio-loud galaxies and in post-burst  
radio-quiet galaxies \citep[e.g.][]{chen10}. This will be the subject of future work.

\section*{acknowledgements}
We are very grateful to P. N. Best for kindly providing his codes for
matching SDSS, NVSS and FIRST.
The research is supported by the National Natural Science Foundation of China (NSFC) under
NSFC-10878010, 10633040, 11003007 and 11133001, the National Basic Research Program (973 program
No. 2007CB815405)  and the National Science Foundation of the United
States Grant No. 0907839.  

Thanks go to the Aspen Center for Physics, which is supported
by  NSF Grant No. 1066293, for hospitality during the writing of this paper.

Funding for SDSS-III has been provided by the Alfred P. Sloan Foundation, the Participating Institutions, the 
National Science Foundation, and the U.S. Department of Energy. SDSS-III is managed by the Astrophysical 
Research Consortium for the Participating Institutions of the SDSS-III Collaboration including the University of 
Arizona, the Brazilian Participation Group, Brookhaven National Laboratory, University of Cambridge, University 
of Florida, the French Participation Group, the German Participation Group, the Instituto de Astrofisica de Canarias, 
the Michigan State/Notre Dame/JINA Participation Group, Johns Hopkins University, Lawrence Berkeley National 
Laboratory, Max Planck Institute for Astrophysics, New Mexico State University, New York University, Ohio State 
University, Pennsylvania State University, University of Portsmouth, Princeton University, the Spanish Participation 
Group, University of Tokyo, University of Utah, Vanderbilt University, University of Virginia, University of Washington, 
and Yale University. 

The research uses the NVSS and FIRST
radio surveys, carried out using the NRAO VLA: NRAO is operated
by Associated Universities Inc., under co-operative agreement
with the National Science Foundation. 

\bibliographystyle{mn2e}

\bibliography{ms}

\begin{thebibliography}{}
\bibitem[\protect\citeauthoryear{{Abazajian}, {Adelman-McCarthy},
  {Ag{\"u}eros}, {Allam}, {Allende Prieto}, {An}, {Anderson}, {Anderson},
  {Annis}, {Bahcall} \& et al.}{{Abazajian} et~al.}{2009}]{abazajian09}
{Abazajian} K.~N.,  {Adelman-McCarthy} J.~K.,  {Ag{\"u}eros} M.~A.,  {Allam}
  S.~S.,  {Allende Prieto} C.,  {An} D.,  {Anderson} K.~S.~J.,  {Anderson}
  S.~F.,  {Annis} J.,  {Bahcall} N.~A.,    et al. 2009, \apjs, 182, 543

\bibitem[\protect\citeauthoryear{{Aihara}, {Allende Prieto}, {An}, {Anderson},
  {Aubourg}, {Balbinot}, {Beers}, {Berlind} \& et al.}{{Aihara}
  et~al.}{2011}]{aihara11}
{Aihara} H.,  {Allende Prieto} C.,  {An} D.,  {Anderson} S.~F.,  {Aubourg}
  {\'E}.,  {Balbinot} E.,  {Beers} T.~C.,  {Berlind} A.~A.,    et al. 2011,
  \apjs, 193, 29


\bibitem[\protect\citeauthoryear{{Becker}, {White} \& {Helfand}}{{Becker}
  et~al.}{1995}]{becker95}
{Becker} R.~H.,  {White} R.~L.,    {Helfand} D.~J.,  1995, \apj, 450, 559

\bibitem[\protect\citeauthoryear{{Best} \& {Heckman}}{{Best} \&
  {Heckman}}{2012}]{best12}
{Best} P.~N.,  {Heckman} T.~M.,  2012, \mnras, 421, 1569

\bibitem[\protect\citeauthoryear{{Best}, {Kaiser}, {Heckman} \&
  {Kauffmann}}{{Best} et~al.}{2006}]{best06}
{Best} P.~N.,  {Kaiser} C.~R.,  {Heckman} T.~M.,    {Kauffmann} G.,  2006,
  \mnras, 368, L67

\bibitem[\protect\citeauthoryear{{Best}, {Kauffmann}, {Heckman} \&
  {Ivezi{\'c}}}{{Best} et~al.}{2005a}]{best05a}
{Best} P.~N.,  {Kauffmann} G.,  {Heckman} T.~M.,    {Ivezi{\'c}} {\v Z}.,
  2005a, \mnras, 362, 9

\bibitem[\protect\citeauthoryear{{Best}, {Kauffmann}, {Heckman}, {Brinchmann},
  {Charlot}, {Ivezi{\'c}} \& {White}}{{Best} et~al.}{2005b}]{best05b}
{Best} P.~N.,  {Kauffmann} G.,  {Heckman} T.~M.,  {Brinchmann} J.,  {Charlot}
  S.,  {Ivezi{\'c}} {\v Z}.,    {White} S.~D.~M.,  2005b, \mnras, 362, 25

\bibitem[\protect\citeauthoryear{{Best}, {von der Linden}, {Kauffmann},
  {Heckman} \& {Kaiser}}{{Best} et~al.}{2007}]{best07}
{Best} P.~N.,  {von der Linden} A.,  {Kauffmann} G.,  {Heckman} T.~M.,
  {Kaiser} C.~R.,  2007, \mnras, 379, 894

\bibitem[\protect\citeauthoryear{{Binney} \& {Tabor}}{{Binney} \&
  {Tabor}}{1995}]{binney95}
{Binney} J.,  {Tabor} G.,  1995, \mnras, 276, 663

\bibitem[\protect\citeauthoryear{{B{\^i}rzan}, {McNamara}, {Nulsen}, {Carilli}
  \& {Wise}}{{B{\^i}rzan} et~al.}{2008}]{birzan08}
{B{\^i}rzan} L.,  {McNamara} B.~R.,  {Nulsen} P.~E.~J.,  {Carilli} C.~L.,
  {Wise} M.~W.,  2008, \apj, 686, 859

\bibitem[\protect\citeauthoryear{{B{\^i}rzan}, {Rafferty}, {McNamara}, {Wise}
  \& {Nulsen}}{{B{\^i}rzan} et~al.}{2004}]{birzan04}
{B{\^i}rzan} L.,  {Rafferty} D.~A.,  {McNamara} B.~R.,  {Wise} M.~W.,
  {Nulsen} P.~E.~J.,  2004, \apj, 607, 800

\bibitem[\protect\citeauthoryear{{Boehringer}, {Voges}, {Fabian}, {Edge} \&
  {Neumann}}{{Boehringer} et~al.}{1993}]{bohringer93}
{Boehringer} H.,  {Voges} W.,  {Fabian} A.~C.,  {Edge} A.~C.,    {Neumann}
  D.~M.,  1993, \mnras, 264, L25

\bibitem[\protect\citeauthoryear{{Bower}, {Benson}, {Malbon}, {Helly}, {Frenk},
  {Baugh}, {Cole} \& {Lacey}}{{Bower} et~al.}{2006}]{bower06}
{Bower} R.~G.,  {Benson} A.~J.,  {Malbon} R.,  {Helly} J.~C.,  {Frenk} C.~S.,
  {Baugh} C.~M.,  {Cole} S.,    {Lacey} C.~G.,  2006, \mnras, 370, 645

\bibitem[\protect\citeauthoryear{{Brinchmann}, {Charlot}, {White}, {Tremonti},
  {Kauffmann}, {Heckman} \& {Brinkmann}}{{Brinchmann}
  et~al.}{2004}]{brinchmann04}
{Brinchmann} J.,  {Charlot} S.,  {White} S.~D.~M.,  {Tremonti} C.,  {Kauffmann}
  G.,  {Heckman} T.,    {Brinkmann} J.,  2004, \mnras, 351, 1151

\bibitem[\protect\citeauthoryear{{Bruzual} \& {Charlot}}{{Bruzual} \&
  {Charlot}}{2003}]{bc03}
{Bruzual} G.,  {Charlot} S.,  2003, \mnras, 344, 1000

\bibitem[\protect\citeauthoryear{{Catinella}, {Schiminovich}, {Kauffmann},
  {Fabello}, {Wang}, {Hummels}, {Lemonias}, {Moran} \& et al.}{{Catinella}
  et~al.}{2010}]{catinella10}
{Catinella} B.,  {Schiminovich} D.,  {Kauffmann} G.,  {Fabello} S.,  {Wang} J.,
   {Hummels} C.,  {Lemonias} J.,  {Moran} S.~M.,    et al. 2010, \mnras, 403,
  683

\bibitem[\protect\citeauthoryear{{Cavagnolo}, {Donahue}, {Voit} \&
  {Sun}}{{Cavagnolo} et~al.}{2008}]{cavagnolo08}
{Cavagnolo} K.~W.,  {Donahue} M.,  {Voit} G.~M.,    {Sun} M.,  2008, \apjl,
  683, L107

\bibitem[\protect\citeauthoryear{{Chen}, {Kauffmann}, {Tremonti}, {White},
  {Heckman}, {Kova{\v c}}, {Bundy}, {Chisholm} \& et al.}{{Chen}
  et~al.}{2012}]{chen12}
{Chen} Y.-M.,  {Kauffmann} G.,  {Tremonti} C.~A.,  {White} S.,  {Heckman}
  T.~M.,  {Kova{\v c}} K.,  {Bundy} K.,  {Chisholm} J.,    et al. 2012, \mnras,
  421, 314

\bibitem[\protect\citeauthoryear{{Chen}, {Tremonti}, {Heckman}, {Kauffmann},
  {Weiner}, {Brinchmann} \& {Wang}}{{Chen} et~al.}{2010}]{chen10}
{Chen} Y.-M.,  {Tremonti} C.~A.,  {Heckman} T.~M.,  {Kauffmann} G.,  {Weiner}
  B.~J.,  {Brinchmann} J.,    {Wang} J.,  2010, \aj, 140, 445

\bibitem[\protect\citeauthoryear{{Chen}, {Wild}, {Kauffmann}, {Blaizot},
  {Davis}, {Noeske}, {Wang} \& {Willmer}}{{Chen} et~al.}{2009}]{chen09}
{Chen} Y.-M.,  {Wild} V.,  {Kauffmann} G.,  {Blaizot} J.,  {Davis} M.,
  {Noeske} K.,  {Wang} J.-M.,    {Willmer} C.,  2009, \mnras, 393, 406

\bibitem[\protect\citeauthoryear{{Churazov}, {Br{\"u}ggen}, {Kaiser},
  {B{\"o}hringer} \& {Forman}}{{Churazov} et~al.}{2001}]{churazov01}
{Churazov} E.,  {Br{\"u}ggen} M.,  {Kaiser} C.~R.,  {B{\"o}hringer} H.,
  {Forman} W.,  2001, \apj, 554, 261

\bibitem[\protect\citeauthoryear{{Colless}, {Dalton}, {Maddox}, {Sutherland},
  {Norberg}, {Cole}, {Bland-Hawthorn}, {Bridges} \& et al.}{{Colless}
  et~al.}{2001}]{colless01}
{Colless} M.,  {Dalton} G.,  {Maddox} S.,  {Sutherland} W.,  {Norberg} P.,
  {Cole} S.,  {Bland-Hawthorn} J.,  {Bridges} T.,    et al. 2001, \mnras, 328,
  1039

\bibitem[\protect\citeauthoryear{{Condon}, {Cotton} \& {Broderick}}{{Condon}
  et~al.}{2002}]{condon02}
{Condon} J.~J.,  {Cotton} W.~D.,    {Broderick} J.~J.,  2002, \aj, 124, 675

\bibitem[\protect\citeauthoryear{{Condon}, {Cotton}, {Greisen}, {Yin},
  {Perley}, {Taylor} \& {Broderick}}{{Condon} et~al.}{1998}]{condon98}
{Condon} J.~J.,  {Cotton} W.~D.,  {Greisen} E.~W.,  {Yin} Q.~F.,  {Perley}
  R.~A.,  {Taylor} G.~B.,    {Broderick} J.~J.,  1998, \aj, 115, 1693

\bibitem[\protect\citeauthoryear{{Croton}, {Springel}, {White}, {De Lucia},
  {Frenk}, {Gao}, {Jenkins}, {Kauffmann} \& et al.}{{Croton}
  et~al.}{2006}]{croton06}
{Croton} D.~J.,  {Springel} V.,  {White} S.~D.~M.,  {De Lucia} G.,  {Frenk}
  C.~S.,  {Gao} L.,  {Jenkins} A.,  {Kauffmann} G.,    et al. 2006, \mnras,
  365, 11

\bibitem[\protect\citeauthoryear{{Donoso}, {Best} \& {Kauffmann}}{{Donoso}
  et~al.}{2009}]{donoso09}
{Donoso} E.,  {Best} P.~N.,    {Kauffmann} G.,  2009, \mnras, 392, 617

\bibitem[\protect\citeauthoryear{{Donoso}, {Li}, {Kauffmann}, {Best} \&
  {Heckman}}{{Donoso} et~al.}{2010}]{donoso10}
{Donoso} E.,  {Li} C.,  {Kauffmann} G.,  {Best} P.~N.,    {Heckman} T.~M.,
  2010, \mnras, 407, 1078

\bibitem[\protect\citeauthoryear{{Dunlop} \& {Peacock}}{{Dunlop} \&
  {Peacock}}{1990}]{dunlop90}
{Dunlop} J.~S.,  {Peacock} J.~A.,  1990, \mnras, 247, 19

\bibitem[\protect\citeauthoryear{{Eisenstein}, {Weinberg}, {Agol}, {Aihara},
  {Allende Prieto}, {Anderson}, {Arns}, {Aubourg}, {Balbinot} \& et
  al.}{{Eisenstein} et~al.}{2011}]{eisenstein11}
{Eisenstein} D.~J.,  {Weinberg} D.~H.,  {Agol} E.,  {Aihara} H.,  {Allende
  Prieto} C.,  {Anderson} S.~F.,  {Arns} J.~A.,  {Aubourg} {\'E}.,  {Balbinot}
  E.,    et al. 2011, \aj, 142, 72

\bibitem[\protect\citeauthoryear{{Fabian}, {Sanders}, {Allen}, {Crawford},
  {Iwasawa}, {Johnstone}, {Schmidt} \& {Taylor}}{{Fabian}
  et~al.}{2003}]{fabian03}
{Fabian} A.~C.,  {Sanders} J.~S.,  {Allen} S.~W.,  {Crawford} C.~S.,  {Iwasawa}
  K.,  {Johnstone} R.~M.,  {Schmidt} R.~W.,    {Taylor} G.~B.,  2003, \mnras,
  344, L43

\bibitem[\protect\citeauthoryear{{Fabian}, {Sanders}, {Ettori}, {Taylor},
  {Allen}, {Crawford}, {Iwasawa}, {Johnstone} \& et al.}{{Fabian}
  et~al.}{2000}]{fabian00}
{Fabian} A.~C.,  {Sanders} J.~S.,  {Ettori} S.,  {Taylor} G.~B.,  {Allen}
  S.~W.,  {Crawford} C.~S.,  {Iwasawa} K.,  {Johnstone} R.~M.,    et al. 2000,
  \mnras, 318, L65

\bibitem[\protect\citeauthoryear{{Fabian}, {Sanders}, {Taylor} \&
  {Allen}}{{Fabian} et~al.}{2005}]{fabian05}
{Fabian} A.~C.,  {Sanders} J.~S.,  {Taylor} G.~B.,    {Allen} S.~W.,  2005,
  \mnras, 360, L20

\bibitem[\protect\citeauthoryear{{Forman}, {Nulsen}, {Heinz}, {Owen}, {Eilek},
  {Vikhlinin}, {Markevitch}, {Kraft} \& et al.}{{Forman}
  et~al.}{2005}]{forman05}
{Forman} W.,  {Nulsen} P.,  {Heinz} S.,  {Owen} F.,  {Eilek} J.,  {Vikhlinin}
  A.,  {Markevitch} M.,  {Kraft} R.,    et al. 2005, \apj, 635, 894

\bibitem[\protect\citeauthoryear{{Fukugita}, {Ichikawa}, {Gunn}, {Doi},
  {Shimasaku} \& {Schneider}}{{Fukugita} et~al.}{1996}]{fukugita96}
{Fukugita} M.,  {Ichikawa} T.,  {Gunn} J.~E.,  {Doi} M.,  {Shimasaku} K.,
  {Schneider} D.~P.,  1996, \aj, 111, 1748

\bibitem[\protect\citeauthoryear{{Gunn}, {Carr}, {Rockosi}, {Sekiguchi},
  {Berry}, {Elms}, {de Haas}, {Ivezi{\'c}} \& et al.}{{Gunn}
  et~al.}{1998}]{gunn98}
{Gunn} J.~E.,  {Carr} M.,  {Rockosi} C.,  {Sekiguchi} M.,  {Berry} K.,  {Elms}
  B.,  {de Haas} E.,  {Ivezi{\'c}} {\v Z}.,    et al. 1998, \aj, 116, 3040

\bibitem[\protect\citeauthoryear{{Gunn}, {Siegmund}, {Mannery}, {Owen}, {Hull},
  {Leger}, {Carey}, {Knapp} \& et al.}{{Gunn} et~al.}{2006}]{gunn06}
{Gunn} J.~E.,  {Siegmund} W.~A.,  {Mannery} E.~J.,  {Owen} R.~E.,  {Hull}
  C.~L.,  {Leger} R.~F.,  {Carey} L.~N.,  {Knapp} G.~R.,    et al. 2006, \aj,
  131, 2332


\bibitem[\protect\citeauthoryear{{Heckman}, {Kauffmann}, {Brinchmann},
  {Charlot}, {Tremonti} \& {White}}{{Heckman} et~al.}{2004}]{heckman04}
{Heckman} T.~M.,  {Kauffmann} G.,  {Brinchmann} J.,  {Charlot} S.,  {Tremonti}
  C.,    {White} S.~D.~M.,  2004, \apj, 613, 109

\bibitem[\protect\citeauthoryear{{Hogg}, {Finkbeiner}, {Schlegel} \&
  {Gunn}}{{Hogg} et~al.}{2001}]{hogg01}
{Hogg} D.~W.,  {Finkbeiner} D.~P.,  {Schlegel} D.~J.,    {Gunn} J.~E.,  2001,
  \aj, 122, 2129


\bibitem[\protect\citeauthoryear{{Johnston}, {Sadler}, {Cannon}, {Croom},
  {Ross} \& {Schneider}}{{Johnston} et~al.}{2008}]{johnston08}
{Johnston} H.~M.,  {Sadler} E.~M.,  {Cannon} R.,  {Croom} S.~M.,  {Ross} N.~P.,
     {Schneider} D.~P.,  2008, \mnras, 384, 692

\bibitem[\protect\citeauthoryear{{Karim}, {Schinnerer},
  {Mart{\'{\i}}nez-Sansigre}, {Sargent}, {van der Wel}, {Rix}, {Ilbert},
  {Smol{\v c}i{\'c}} \& et al.}{{Karim} et~al.}{2011}]{karim11}
{Karim} A.,  {Schinnerer} E.,  {Mart{\'{\i}}nez-Sansigre} A.,  {Sargent} M.~T.,
   {van der Wel} A.,  {Rix} H.-W.,  {Ilbert} O.,  {Smol{\v c}i{\'c}} V.,    et
  al. 2011, \apj, 730, 61

\bibitem[\protect\citeauthoryear{{Kauffmann}, {Heckman} \& {Best}}{{Kauffmann}
  et~al.}{2008}]{kauffmann08}
{Kauffmann} G.,  {Heckman} T.~M.,    {Best} P.~N.,  2008, \mnras, 384, 953

\bibitem[\protect\citeauthoryear{{Kauffmann}, {Heckman}, {White}, {Charlot},
  {Tremonti}, {Brinchmann}, {Bruzual}, {Peng} \& et al.}{{Kauffmann}
  et~al.}{2003a}]{kauffmann03a}
{Kauffmann} G.,  {Heckman} T.~M.,  {White} S.~D.~M.,  {Charlot} S.,  {Tremonti}
  C.,  {Brinchmann} J.,  {Bruzual} G.,  {Peng} E.~W.,    et al. 2003a, \mnras,
  341, 33

\bibitem[\protect\citeauthoryear{{Kauffmann}, {Heckman}, {White}, {Charlot},
  {Tremonti}, {Peng}, {Seibert}, {Brinkmann} \& et al.}{{Kauffmann}
  et~al.}{2003b}]{kauffmann03b}
{Kauffmann} G.,  {Heckman} T.~M.,  {White} S.~D.~M.,  {Charlot} S.,  {Tremonti}
  C.,  {Peng} E.~W.,  {Seibert} M.,  {Brinkmann} J.,    et al. 2003b, \mnras,
  341, 54

\bibitem[\protect\citeauthoryear{{Kauffmann}, {Heckman}, {Tremonti},
  {Brinchmann}, {Charlot}, {White}, {Ridgway}, {Brinkmann} \& et
  al.}{{Kauffmann} et~al.}{2003c}]{kauffmann03c}
{Kauffmann} G.,  {Heckman} T.~M.,  {Tremonti} C.,  {Brinchmann} J.,  {Charlot}
  S.,  {White} S.~D.~M.,  {Ridgway} S.~E.,  {Brinkmann} J.,    et al. 2003c,
  \mnras, 346, 1055

\bibitem[\protect\citeauthoryear{{Mandelbaum}, {Li}, {Kauffmann} \&
  {White}}{{Mandelbaum} et~al.}{2009}]{mandelbaum09}
{Mandelbaum} R.,  {Li} C.,  {Kauffmann} G.,    {White} S.~D.~M.,  2009, \mnras,
  393, 377

\bibitem[\protect\citeauthoryear{{Mauch} \& {Sadler}}{{Mauch} \&
  {Sadler}}{2007}]{mauch07}
{Mauch} T.,  {Sadler} E.~M.,  2007, \mnras, 375, 931

\bibitem[\protect\citeauthoryear{{Padmanabhan}, {Schlegel}, {Finkbeiner},
  {Barentine}, {Blanton}, {Brewington}, {Gunn}, {Harvanek} \& et
  al.}{{Padmanabhan} et~al.}{2008}]{pad08}
{Padmanabhan} N.,  {Schlegel} D.~J.,  {Finkbeiner} D.~P.,  {Barentine} J.~C.,
  {Blanton} M.~R.,  {Brewington} H.~J.,  {Gunn} J.~E.,  {Harvanek} M.,    et
  al. 2008, \apj, 674, 1217


\bibitem[\protect\citeauthoryear{{Pier}, {Munn}, {Hindsley}, {Hennessy},
  {Kent}, {Lupton} \& {Ivezi{\'c}}}{{Pier} et~al.}{2003}]{pier03}
{Pier} J.~R.,  {Munn} J.~A.,  {Hindsley} R.~B.,  {Hennessy} G.~S.,  {Kent}
  S.~M.,  {Lupton} R.~H.,    {Ivezi{\'c}} {\v Z}.,  2003, \aj, 125, 1559



\bibitem[\protect\citeauthoryear{{Rawlings} \& {Jarvis}}{{Rawlings} \&
  {Jarvis}}{2004}]{rawlings04}
{Rawlings} S.,  {Jarvis} M.~J.,  2004, \mnras, 355, L9

\bibitem[\protect\citeauthoryear{{Rola}, {Terlevich} \& {Terlevich}}{{Rola}
  et~al.}{1997}]{rola97}
{Rola} C.~S.,  {Terlevich} E.,    {Terlevich} R.~J.,  1997, \mnras, 289, 419

\bibitem[\protect\citeauthoryear{{Sadler}, {Jackson}, {Cannon}, {McIntyre},
  {Murphy}, {Bland-Hawthorn}, {Bridges}, {Cole} \& et al.}{{Sadler}
  et~al.}{2002}]{sadler02}
{Sadler} E.~M.,  {Jackson} C.~A.,  {Cannon} R.~D.,  {McIntyre} V.~J.,  {Murphy}
  T.,  {Bland-Hawthorn} J.,  {Bridges} T.,  {Cole} S.,    et al. 2002, \mnras,
  329, 227

\bibitem[\protect\citeauthoryear{{Saintonge}, {Kauffmann}, {Kramer}, {Tacconi},
  {Buchbender}, {Catinella}, {Fabello}, {Graci{\'a}-Carpio} \& et
  al.}{{Saintonge} et~al.}{2011}]{saintonge11}
{Saintonge} A.,  {Kauffmann} G.,  {Kramer} C.,  {Tacconi} L.~J.,  {Buchbender}
  C.,  {Catinella} B.,  {Fabello} S.,  {Graci{\'a}-Carpio} J.,    et al. 2011,
  \mnras, 415, 32

\bibitem[\protect\citeauthoryear{{Salim}, {Fang}, {Rich}, {Faber} \&
  {Thilker}}{{Salim} et~al.}{2012}]{salim12}
{Salim} S.,  {Fang} J.~J.,  {Rich} R.~M.,  {Faber} S.~M.,    {Thilker} D.~A.,
  2012, ArXiv e-prints

\bibitem[\protect\citeauthoryear{{Schawinski}, {Khochfar}, {Kaviraj}, {Yi},
  {Boselli}, {Barlow}, {Conrow}, {Forster} \& et al.}{{Schawinski}
  et~al.}{2006}]{schawinski06}
{Schawinski} K.,  {Khochfar} S.,  {Kaviraj} S.,  {Yi} S.~K.,  {Boselli} A.,
  {Barlow} T.,  {Conrow} T.,  {Forster} K.,    et al. 2006, \nat, 442, 888

\bibitem[\protect\citeauthoryear{{Schawinski}, {Lintott}, {Thomas}, {Kaviraj},
  {Viti}, {Silk}, {Maraston}, {Sarzi} \& et al.}{{Schawinski}
  et~al.}{2009}]{schawinski09}
{Schawinski} K.,  {Lintott} C.~J.,  {Thomas} D.,  {Kaviraj} S.,  {Viti} S.,
  {Silk} J.,  {Maraston} C.,  {Sarzi} M.,    et al. 2009, \apj, 690, 1672

\bibitem[\protect\citeauthoryear{{Schlegel}, {Finkbeiner} \&
  {Davis}}{{Schlegel} et~al.}{1998}]{schlegel98}
{Schlegel} D.~J.,  {Finkbeiner} D.~P.,    {Davis} M.,  1998, \apj, 500, 525

\bibitem[\protect\citeauthoryear{{Strauss}, {Weinberg}, {Lupton}, {Narayanan},
  {Annis}, {Bernardi}, {Blanton}, {Burles} \& et al.}{{Strauss}
  et~al.}{2002}]{strauss02}
{Strauss} M.~A.,  {Weinberg} D.~H.,  {Lupton} R.~H.,  {Narayanan} V.~K.,
  {Annis} J.,  {Bernardi} M.,  {Blanton} M.,  {Burles} S.,    et al. 2002, \aj,
  124, 1810

\bibitem[\protect\citeauthoryear{{Tremonti}, {Heckman}, {Kauffmann},
  {Brinchmann}, {Charlot}, {White}, {Seibert}, {Peng} \& et al.}{{Tremonti}
  et~al.}{2004}]{tremonti04}
{Tremonti} C.~A.,  {Heckman} T.~M.,  {Kauffmann} G.,  {Brinchmann} J.,
  {Charlot} S.,  {White} S.~D.~M.,  {Seibert} M.,  {Peng} E.~W.,    et al.
  2004, \apj, 613, 898

\bibitem[\protect\citeauthoryear{{Wake}, {Sheth}, {Nichol}, {Baugh},
  {Bland-Hawthorn}, {Colless}, {Couch}, {Croom} \& et al.}{{Wake}
  et~al.}{2008a}]{wake08a}
{Wake} D.~A.,  {Sheth} R.~K.,  {Nichol} R.~C.,  {Baugh} C.~M.,
  {Bland-Hawthorn} J.,  {Colless} M.,  {Couch} W.~J.,  {Croom} S.~M.,    et al.
  2008a, \mnras, 387, 1045


\bibitem[\protect\citeauthoryear{{Wake}, {Croom}, {Sadler} \&
  {Johnston}}{{Wake} et~al.}{2008b}]{wake08b}
{Wake} D.~A.,  {Croom} S.~M.,  {Sadler} E.~M.,    {Johnston} H.~M.,  2008b,
  \mnras, 391, 1674

\bibitem[\protect\citeauthoryear{{White} \& {Frenk}}{{White} \&
  {Frenk}}{1991}]{white91}
{White} S.~D.~M.,  {Frenk} C.~S.,  1991, \apj, 379, 52

\bibitem[\protect\citeauthoryear{{York}, {Adelman}, {Anderson} Jr., {Anderson},
  {Annis}, {Bahcall}, {Bakken}, {Barkhouser} \& et al.}{{York}
  et~al.}{2000}]{york00}
{York} D.~G.,  {Adelman} J.,  {Anderson} Jr. J.~E.,  {Anderson} S.~F.,  {Annis}
  J.,  {Bahcall} N.~A.,  {Bakken} J.~A.,  {Barkhouser} R.,    et al. 2000, \aj,
  120, 1579

\bibitem[\protect\citeauthoryear{{Zheng}, {Bell}, {Papovich}, {Wolf},
  {Meisenheimer}, {Rix}, {Rieke} \& {Somerville}}{{Zheng}
  et~al.}{2007}]{zheng07}
{Zheng} X.~Z.,  {Bell} E.~F.,  {Papovich} C.,  {Wolf} C.,  {Meisenheimer} K.,
  {Rix} H.-W.,  {Rieke} G.~H.,    {Somerville} R.,  2007, \apjl, 661, L41

\end{thebibliography}
\end{document}